\documentclass{aa}  
\usepackage{txfonts,textcomp}
\makeatletter
\renewcommand*\aa@pageof{, page \thepage{} of \pageref*{LastPage}}
\makeatother
\begin{document} 
\nolinenumbers
\title{Forgotten treasures in the HST/FOC UV imaging polarimetric archives of active galactic nuclei}
\subtitle{III. Five years monitoring of M87}
\author{F. Marin\inst{1}\thanks{\href{mailto:frederic.marin@astro.unistra.fr}{frederic.marin@astro.unistra.fr}}
    \and
    T. Barnouin\inst{1}
    \and
    K. Wu\inst{2,3}
    \and
    E. Lopez-Rodriguez\inst{4}
}
\institute{Universit\'e de Strasbourg, CNRS, Observatoire astronomique de Strasbourg, UMR 7550, F-67000 Strasbourg, France
    \and
    Mullard Space Science Laboratory, University College London, Holmbury St Mary, Surrey RH5 6NT, UK
    \and 
    Kavli Institute for the Physics and Mathematics of the Universe (WPI), UTIAS, The University of Tokyo, Kashiwa, Chiba 277-8583, Japan
    \and
    Kavli Institute for Particle Astrophysics \& Cosmology (KIPAC), Stanford University, Stanford, CA 94305, USA     
}
\date{Received July 5, 2024; accepted October 8, 2024}
%
%
\abstract
{The active galactic nucleus (AGN) within M87, a giant elliptical galaxy, is responsible for one of the closest kiloparsec-scale relativistic jet to Earth. It is thus a perfect target for spatially resolved observations.}
{This one-sided jet has been extensively observed at almost all wavelengths, with almost all techniques. Among many other discoveries, it was found that the optical emission is more concentrated in the knots and along the center line of the jet, in comparison to, e.g., radio emission. But what can we learn from its polarized counterpart?}
{We unearthed unpublished polarization maps taken with the Faint Object Camera (FOC) aboard the Hubble Space Telescope (HST), obtained between 1995 and 1999. At a rate of one observation per year, we can follow the evolution of the polarized flux knots in the jet. We can thus constrain the time scale of variation of the magnetic field up to a spatial resolution of one tenth of an arcsecond ($\sim$ 11.5~pc).}
{After coherently reducing the five observations using the same methodology presented in the first paper of this series, the analysis of polarized maps from POS~1 (base of the jet) and POS~3 (end of the jet) reveals significant temporal and spatial dynamics in the jet's magnetic fields morphology. Despite minimal changes in overall intensity structure, notable fluctuations in polarization degrees and angles are detected across various knots and inter-knot regions. In addition, the emission and polarization characteristics of M87's jet differ significantly between POS1 and POS3. POS1 shows a more collimated jet with strong variability in polarization, while POS3 reveals a thicker structure, a quasi-absence of variability and complex magnetic field interactions. This suggests that the jet may have co-axial structures with distinct kinetic properties. Theoretical models like the jet-in-jet scenario, featuring double helical magnetic flux ropes, help explain these observations, indicating a strong density contrast and higher speeds in the inner jet.}
{Our temporal analysis demonstrates the importance of high spatial resolution polarization mapping in understanding jets polarization properties and overall dynamics, especially if such maps are taken at different wavelengths (ultraviolet and radio hereby).}
\maketitle
\nolinenumbers

%
\section{Introduction}

Situated at a Hubble distance of 23.79~Mpc $\pm$ 1.70~Mpc ($z \approx 0.00428$, \citealt{Cappellari2011}) lies M87, also known as Virgo A, NGC~4486 or 3C~274. It is a giant elliptical E$_0$ galaxy near the center of the Virgo Cluster that harbors a low-luminosity active galactic nucleus (AGN) and a prominent, non-thermal, one-sided jet. The AGN at the heart of M87 has a bolometric luminosity of the order of a few $10^{42}$~erg~s$^{-1}$, but it exhibits flux variations on both short (day) and long time-scales (month/year), with more significant fluctuations at high energies (X-rays) than at low energies (infrared), see, e.g., \citet{Prieto2016} and \citet{Cheng2023}. Such variations, likely related to the accretion mode of the disk surrounding the central black hole, coupled to its low-luminosity and estimated mass (ranging from one to almost ten billion solar masses, see \citealt{Walsh2013,EHT2019,Liepold2023,Simon2024}) classifies M87's accretion mechanism as an Advection Dominated Accretion Flow (ADAF, \citealt{Narayan1995}). This source is thus a very good candidate to understand ADAFs (or alternative models for accretion-ejection, such as  magnetically dominated ergomagnetospheres, \citealt{Blandford2022}) and how they can lead to kiloparsec-scale collimated jets \citep{Nagar2005,Cruz-Osorio2022,Lu2023}.

Indeed, if M87 is famous, it is mostly for its jet(s). First detected in optical by \citet{Curtis1918}, well before the advent of radio-astronomy, this AGN exhibits two giant radio lobes in the North-West and South-East directions. The North-West lobe is connected to the core of M87 by a $\sim 25$~arcseconds-long collimated jet at position angle $\sim 288^\circ$ \citep{Nalewajko2020}. The viewing angle of this jet is estimated to be $10^\circ - 20^\circ$ \citep{Biretta1999,Walker2018}. The counter-jet, connecting the core to the South-East lobe, is undetected but the presence of optical synchrotron emission at the location of the radio peak in the lobe has been revealed by imaging polarization, providing solid proof of the existence of this counter-jet \citep{Sparks1992,Stiavelli1992}. Focusing on the detected jet structure, radio and optical images revealed that the jet emission is not uniform. It rather shows an alternation of bright knots and relatively darker inter-knots regions in the optical and radio bands \citep{Owen1980}. On large scales, the optical and radio jet morphologies appear to be similar, but significant systematic differences pop up at smaller scales. \cite{Sparks1996} have shown that the optical emission is more concentrated in the knots and along the center line of the jet (the jet appearing narrower than in the radio). In addition, high resolution radio data showed that the emission from the knots has a filamentary structure that is better explained if the emission is essentially produced in a boundary layer between the jet and the external medium \citep{Owen1989,Pasetto2021}. Such finding would support the model proposed by \cite{Tavecchio2008}, where the jet is composed of an inner, relativistic spine and of a slower moving, outer sheath responsible for low-frequency emission.

Observing such jet with polarimetry allows to reveal the magnetic field lines. Indeed, the electric vector position angle measured by polarimetry is perpendicular to the magnetic field direction and thus traces the projected magnetic topology of the jet (see, e.g., \citealt{Capetti1997}). Among other results brought by spatially-resolved polarimetry, it was found that the magnetic topology is rather complex, with variations observed at all physical scales \citep{Owen1989}. The degree of polarization is also spatially-dependent, reaching up to $60\%$ in the ultraviolet band \citep{Capetti1997}. This waveband is of particular interest since ultraviolet photons produced by synchrotron emission have a synchrotron lifetime far shorter than the one produced in the radio. Ultraviolet photons are produced in the regions close to where the electrons are accelerated and therefore provide a much cleaner view of the inner jet regions than at longer wavelengths. In addition, the light of old stellar populations from the host is greatly diminished in comparison to observations made in the optical or infrared bands, allowing a cleaner view of the jet structure. 

The only known (and published) ultraviolet polarization map of M87 jet at sub-arcsecond resolution was obtained with the Faint Object Camera (FOC) aboard the Hubble Space Telescope (HST) and dates from 1993 June 17, before the Refurbishment Mission. The data presented by \citet{Capetti1997}, despite suffering from the primary mirror’s flaw, lead to several important conclusions. It was found that the polarization degree is, on average, of the order of $30\%$ over most of the jet, with variations up to $60\%$ thanks to highly ordered magnetic fields and down to $10\%$ due to unresolved small scale structures that have a cancelling effect onto the observed polarization. Finally, no significant depolarization or Faraday rotation have been detected. But how does it hold after the installation of the corrective optics on HST? More importantly, does the jet show temporal variation in flux and polarized flux? If so, are they correlated?

To answer those questions, we searched the FOC archives and discovered that a series of five polarimetric observations had been acquired between 1995 and 1999, at the rate of one observation per year. These data, presented in total flux in \citet{Biretta1999}, have never been analyzed nor published in polarized flux. As part of our FOC data homogenization project, we therefore took the initiative of recovering these observations and reducing them to extract their polarimetric information. The observations themselves and the data reduction process are presented in Sec.~\ref{Observation}. The analysis of the first 7~arcseconds ($\sim$ 644~pc) of the jet, called POS~1 (following the HST nomenclature), is developed in Sect.~\ref{POS~1}, while the last 7~arcseconds of the jet, called POS~3, are explored in Sect.~\ref{POS~3}. Data for POS~2 pertaining to the central part of the jet are not available in the HST archive. We discuss our results in Sect.~\ref{Discussion} and conclude this work in Sect.~\ref{Conclusion}.

\section{Observation and data reduction}
\label{Observation}

The M87 jet observations were taken by the FOC aboard HST from 1995 to 1999, at the rate of one observation per year (see Tab.~\ref{Tab:Obs}). Observations were made in the f/96 mode ($512 \times 512$ pixels format) with the zoom off, providing a spatial resolution of $0.01435 \times 0.01435$~arcsecond$^2$ per pixel, which corresponds to an effective resolution of $35$ milliarcseconds for a $7 \times 7$~arcseconds$^2$ field of view (FoV). Exposures were made in the F342W broadband ultraviolet filter centered on 3420~\AA, in addition to the three polarizing Rochon prisms (POL0, POL60, POL120). Approximately 1.6~ks were accumulated per polarizer per year, leading to a total exposure time of $\sim 1.3$~hours per observation in polarimetric mode. \citet{Biretta1999} report that "several measures were taken to optimize the geometric stability and repeatability of the images", such as FOC warming up, preliminary shots to carefully place the jet on the same region of the detector and internal flats to remove the reseau marks (see \citealt{Biretta1999} and \citealt{Nota1996} for further details).

\begin{table} 
\centering
\begin{tabular}{c c c c}
\hline\hline
\textbf{Obs. ID} & \textbf{Object} & \textbf{Date (year-month-day)} & \textbf{Date (MJD)} \\ 
\hline
5941 & POS~1 & 1995 Jul 5 & 2449903 \\
5941 & POS~3 & 1995 Jul 9 & 2449907 \\
6775 & POS~1 & 1996 Jul 31 & 2450295 \\
6775 & POS~3 & 1996 Aug 5 & 2450300 \\
7274 & POS~1 & 1997 Jul 14 & 2450643 \\
7274 & POS~3 & 1997 Aug 1 & 2450661 \\
7274 & POS~1 & 1998 Jul 20 & 2451014 \\
7274 & POS~3 & 1998 Jul 31 & 2451025 \\
7274 & POS~1 & 1999 Jun 11 & 2451340 \\
7274 & POS~3 & 1999 Jun 17 & 2451346 \\
\hline
\end{tabular}
\caption{Observation log of the HST/FOC dataset for the M87 jet.}   
\label{Tab:Obs} 
\end{table}

To reduce the data downloaded from the MAST HST Legacy Archive\footnote{\url{https://archive.stsci.edu/missions-and-data/hst}}, we followed the guideline and used the automatized, generalized reduction pipeline presented in \citet{Barnouin2023}, the first paper of this series. All necessary information are provided in this technical paper. Here, we only describe the parametrization used to obtain the fully reduced polarization maps. After downloading the data, the raw POL0, POL60 and POL120 images were cropped to show the same region of interest. Background estimation was achieved by looking at the intensity histogram of each image using the Freedman-Diaconis sampling rule. We fitted a Gaussian on the histogram and selected the mean value as representative of the background level. Images were aligned to a precision of $\sim 0.1$ pixels, then the data were binned at $0.1 \times 0.1$~arcseconds$^2$ to maximize both the spatial resolution and the signal-to-noise ratio. The images were smoothed using a Gaussian kernel with a full-width at half-maximum of 0.2~arcseconds, i.e. twice the size of a resampled pixel. For each pixel, the Stokes parameters I, Q and U were computed (with their associated uncertainties), as well as the debiased polarization degree $P$ and the electric vector position angle $\Psi$. The final maps were rotated to have the North up ($\Psi = 0^\circ$). The rotation of $\Psi$ follows the IAU convention (the value for the electric-vector position angle of polarization starts from North and increases through East.). No deconvolution was applied during the process. 

\begin{figure*}
\centering
\includegraphics[width=.49\textwidth]{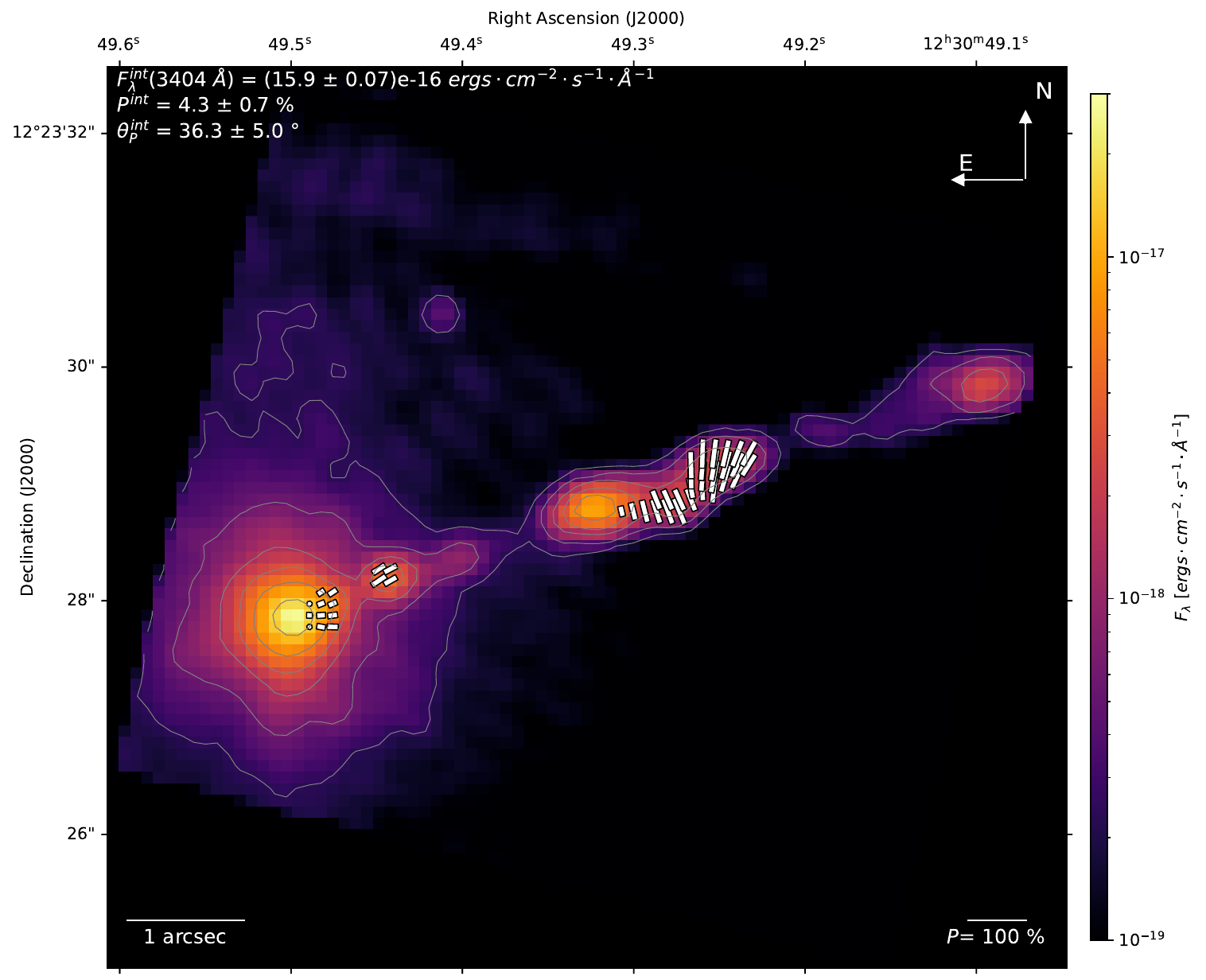}
\includegraphics[width=.49\textwidth]{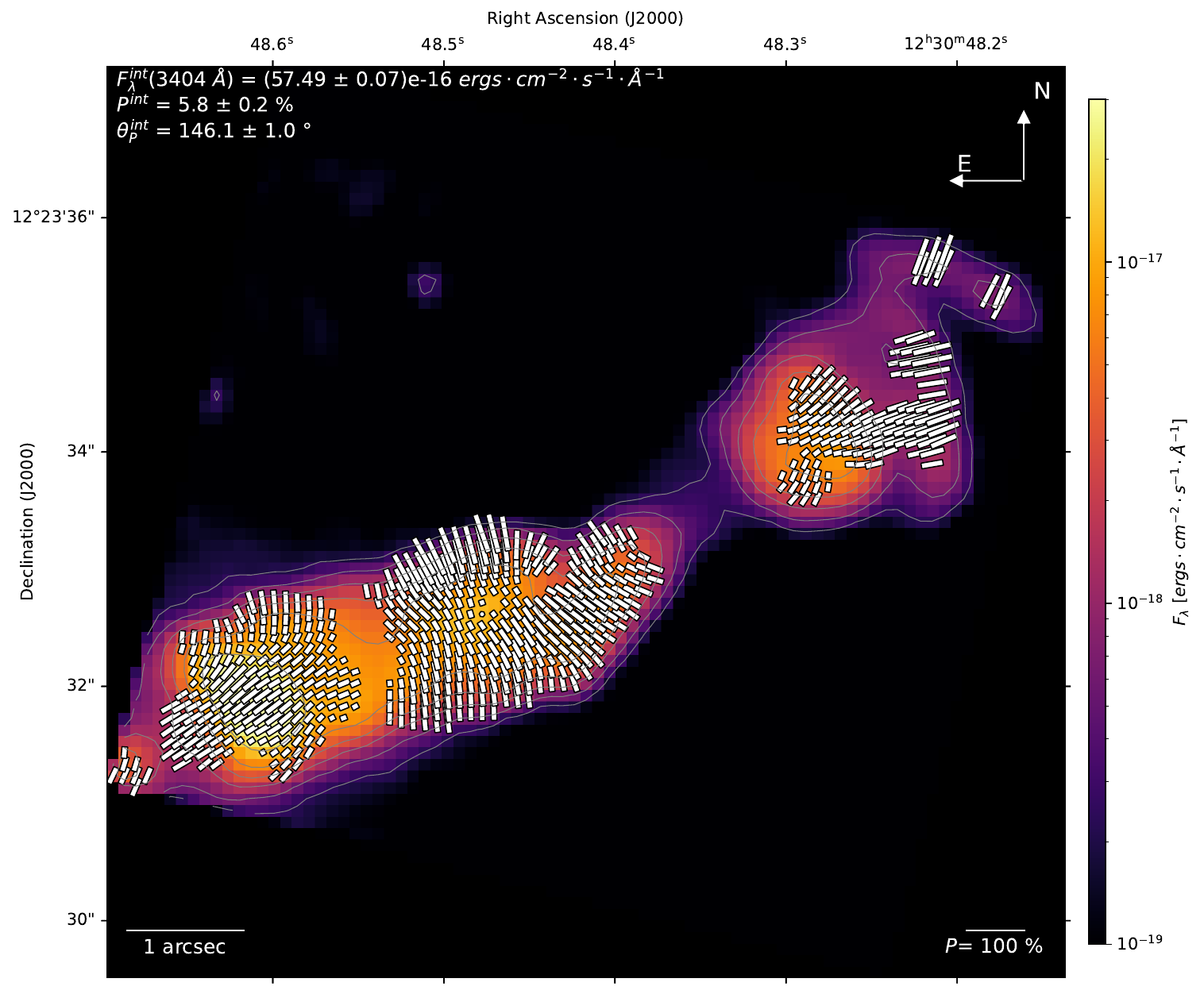}
\caption{Data products obtained from our reduction pipeline. Left : the first 7~arcseconds of the jet (POS~1, 1995). Right : the last 7~arcseconds of the jet (POS~3, 1995). In both cases, the total flux (in ergs~cm$^{-2}$~s$^{-1}$~\AA$^{-1}$) is color-coded and the flux contours are displayed for 0.8\%, 2\%, 5\%, 10\%, 20\% and 50\% of the maximum flux. The polarization information is superimposed to the image using white vectors. $P$ is proportional to the vector length and the angle of $\Psi$ is given by the orientation of the vector. See Sect.~\ref{POS~1} (POS~1) and Sect.~\ref{POS~3} (POS~3) for an analysis of the images. Full scale images with increased visibility of the polarization vectors are presented in appendices.}
\label{Fig:data_reduction}
\end{figure*}

In the final images obtained for POS~1 and POS~3 over the five epochs (see Fig.\ref{Fig:data_reduction} for two examples), the M87 core was found to be slightly count-rate saturated, but it concerns only a single pixel at the heart of the AGN. This pixel was not accounted for when we integrated the polarized signal over the AGN. For all the maps presented in this paper, we only show the polarization information in bins were $P$ is detected at a significance level equal or superior to 3. Finally, at a distance of 23.79~Mpc, 0.1~arcseconds (one spatial bin) corresponds to a linear size of $\sim$ 11.5~pc.

\section{Analysis : POS~1}
\label{POS~1}

\subsection{An absence of morphological changes}
\label{POS~1:Morphology}

\begin{figure*}
\centering
\includegraphics[trim={0 3cm 0 0},clip, width=0.9\textwidth]{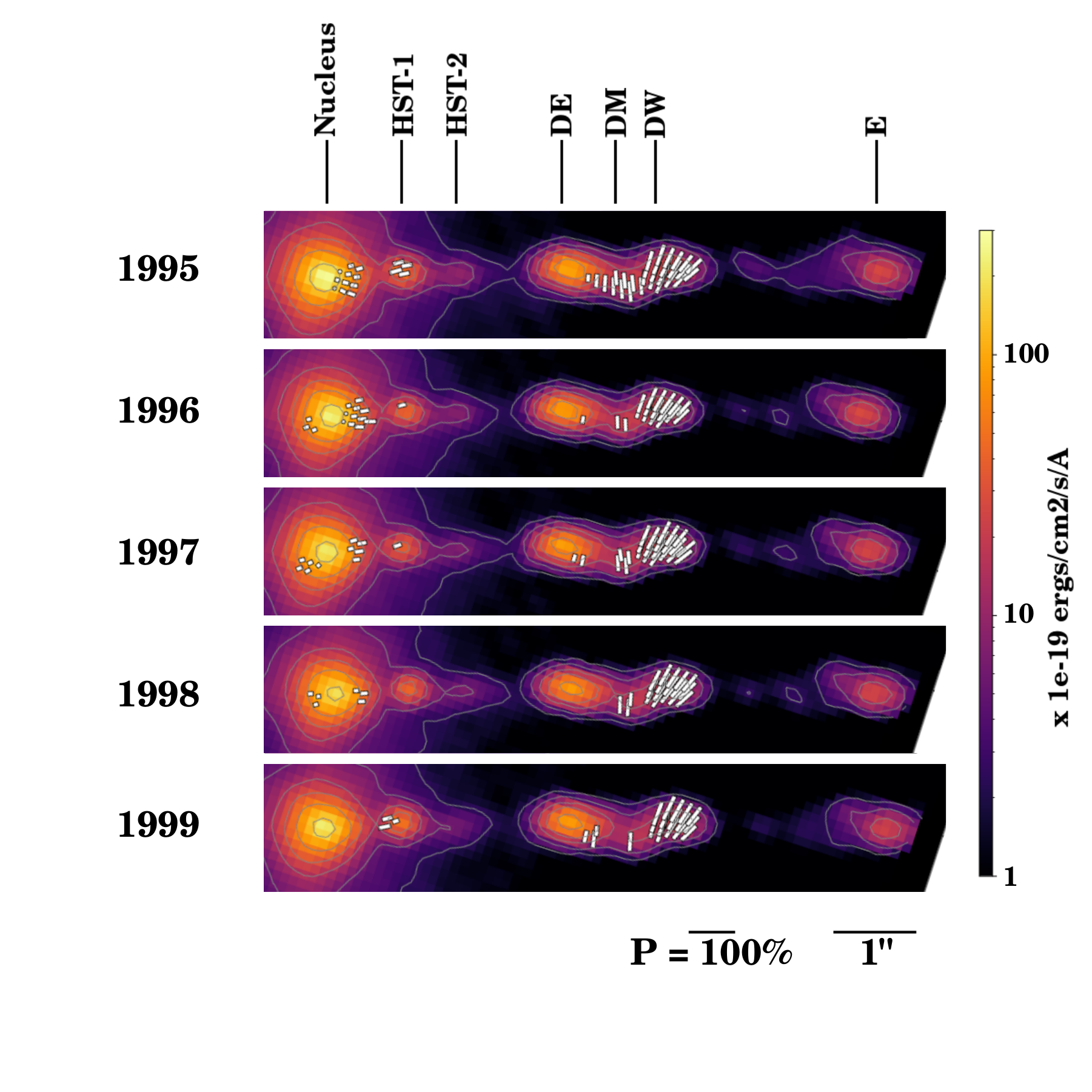}
\caption{Crops of the POS~1 polarized images of the jet polarization from the five epochs examined in this paper (from 1995, top, to 1999, bottom). The images have been rotated from their normal appearances on the sky by $18^\circ$, in such a way that the horizontal axis is along a position angle of $\sim 270^\circ$. The location of the main jet features are marked in black, following the nomenclature of \citet{Biretta1999}. The total flux is color-coded and the polarization is encrypted in the same way as in Fig.~\ref{Fig:data_reduction}.}
\label{Fig:POS~1_frise}
\end{figure*}

We start by analyzing the five POS~1 polarized maps obtained with the FOC. In order not to overload the article with figures, all the POS~1 reduced images are presented in full scale in Appendix~\ref{AppendixA}. Here, we rather summarize the five maps using a sequence of F342W polarized images showing the first 7~arcseconds of the jet in M87 between 1995 and 1999. It is presented in Fig.\ref{Fig:POS~1_frise}. The brightest knots (in total flux) are labeled similarly to \citet{Biretta1999}. At first glance, there seems to be little-to-no variation in morphology of the main jet features. The only variation which is clearly visible is the decrease in total flux of the HST-2 region and of the inter-knot region between knots DW and E (4.5-5.5 arcseconds from the nucleus) that gets fainter by a factor $\sim$ 1.5 and less sharp from 1995 to 1999. However, the shape of the knots, revealed in total flux, does not change in five years at the spatial resolution presented here.

The absence of morphological change is logical, as the proper motions measured by the Karl G. Jansky Very Large Array (VLA) is up to 3$c$ (or 0.012~arcseconds per year) for, e.g., a bright feature in knot D, which is situated 3~arcseconds away from the nucleus \citep{Biretta1995}. At the spatial resolution presented in the polarized images (0.1~arcsecond), it would take 9 HST cycles (9 years) to detect any morphological change. Superluminal motion was indeed detected in the M87 jet by \citet{Biretta1999} with the same data
when examined at a spatial resolution ten times smaller, coincident with the native resolution of the instrument. For safety's sake, we made sure that we also find these superluminal features when we resume to the native spatial resolution of our maps (i.e. 0.01435 $\times$ 0.01435~arcsecond$^2$ bins). However, at these spatial scales, the signal-to-noise in polarimetry is far too weak to extract viable information, which is why we increased the spatial scale of our images up to 0.1~arcsecond.

\subsection{Variability of the source components}
\label{POS~1:Variability}

Although it is impossible to directly detect any morphological changes in the jet with our spatial binning, the polarization vectors superimposed to the bright features in Fig. \ref{Fig:POS~1_frise} are found to vary with time. This is particularly visible in the knots HST-1, DE and DM. Polarization is also detected at more than 3 sigmas at the base of the jet, near the core, both in the direction of the jet and of the counter-jet. The peak of polarization emission appears to be around 1996. After that, the number of pixels with a significant polarization measurement at the jet/counter-jet bases diminishes with time, up to a complete disappearance in 1999. This could very well be due to shocks that enhanced the observed polarization in 1995-1996 at a spatial scale way smaller than what is resolved here. The high degree of polarization detected at this location ($\sim 8\%$ in 1996) with a polarization angle parallel to the jet are tell-tail signatures of such events \citep{Marscher1985,Liodakis2022}. 

To better characterize the variation of $P$ and $\Psi$ with time, we integrated the total and polarized fluxes of each of the known knots and reported their value in Tab.~\ref{Tab:Knots}. The aperture radii used for each of the features are the one proposed by \citet{Biretta1999}, see their Tab.~2. We can see that, when integrating the whole FoV, the total flux gradually diminishes by a factor 1.22 in five years, while the polarization degree shows more chaotic variations, varying on a scale from 0.2\% and 5\%. The polarization angle rotates stochastically from -29.4$^\circ$ to 41.3$^\circ$. Such behavior is not surprising as M87 is known to be the site of powerful variability events \citep{Perlman2011}. The absence of correlation between the flux and $P$ indicates that the overall polarization from the source is, as expected, the sum of contributions from its nucleus and jet features.

The nucleus itself is brighter by a factor 12 -- 100 with respect to the jet knots. It shows a similar decrease in flux with time with respect to the integrated FoV, indicating that the total flux of the source is dominated by the AGN core. Its polarization degree does not follow the same trend and appear to be stochastic at best, with a peak of $P$ in 1996. Nevertheless, the observed $\Psi$ stays approximately parallel to the jet axis at all time, although slight time-dependent changes of the angle values are reported ($\le$ 14$^\circ$ from the average value). The observed polarization (both in $P$ and $\Psi$) is consistent with the values reported by \citet{Fraix1989} in the U band, and slightly larger than what is reported by \citet{Fresco2020} in the the V and I bands ($2 - 3\%$). The difference of $P$ between the ultraviolet and optical bands is simply due to the reduced fraction of host starlight dilution at shorter wavelengths.

HST-1, which is a knot known for its historical increase in optical/ultraviolet brightness by a factor of more than a hundred between 2000 and 2005, has been extensively studied. \citet{Perlman2011} achieved optical polarimetry of this region and found a highly significant correlation between the total flux and $P$, with $P$ varying from 20\% to 40\%, while the orientation of $\Psi$ stays constant. During the HST/FOC five years of observations, HST-1 is found to be variable both in flux and in polarization. The total flux is maximum in 1995, decreases from 1995 to 1997, where it is at its dimmest value, and then rises again until 1999. $P$ varies from $\sim$ 12\% to $\sim$ 23\% but without obvious correlation with the total flux. During the five years monitoring, $\Psi$ can be considered as constant (within the measured error bars) and oriented along the jet axis, as is \citealt{Perlman2011}. It is interesting to note that the sudden increase of HST-1 luminosity that started in 2000 may have shown an early sign of activity in 1999.

The HST-2 knot is much dimmer than HST-1 and shows no specific trend in total flux (maximum in 1995, getting dimmer until 1997, rises in 1998 and then becomes very weak in 1999). Consequently, the measured polarization degree is not statistically significant and only upper limits can be derived. In HST-2, $P$ is likely lower than 20\% over the five years of observation.

The DE and DM fluxes are decreasing with time (slightly for DE, more strongly for DM) while the flux of DW is apparently constant between 1995 and 1997, and then starts to decrease in 1998-1999. For all three knots, there is no significant changes in their polarization degrees and angles over time, with the exception of a rise in $P$ for DE in 1997. Knots DM and DE show the strongest $P$ in POS~1, of the order of $30 - 50\%$, and the DW knot appears to have a polarization angle different to that of DE and DM. It is evident from the figures in Appendix~\ref{AppendixA} that the magnetic field lines are changing their topology within knot DW. DW is clearly a knot with its own magnetic properties.

Finally, knot E appears to be also decreasing in flux with time, while its polarization degree varies from $\sim 10\%$ to $\sim 23\%$. Its polarization angle is distinctly different in 1995 than in 1996-1999, a variation that is possibly linked with the strong dimming of the inter-knot region right before E. The rotation of $\Psi$, the fluctuation of $P$ and the dimming of the inter-knot region before E could possibly indicate the evanescent signatures of an old shock.

From this analysis, it is clear that the nucleus drives the global flux and polarization properties of M87 when the signal is integrated over the whole FoV. However, each cloud shows distinctive signatures in flux (dimming, re-brightening, constant values) and there is no clear correlations between the total flux and $P$, nor between $P$ and $\Psi$. The local variations observed within the knots on the polarized images are likely originating from shocks and magnetic reconnections below the map resolution (0.1~arcsecond).

\begin{table*}
\centering
\begin{tabular}{l c c c c}
\hline\hline
\textbf{Region} & \textbf{Date} & \textbf{Total flux} & \textbf{$P$} & \textbf{$\Psi$} \\ 
\textbf{(aperture in arcsec)} & \textbf{(MJD)} & \textbf{(ergs~cm$^{-2}$~s$^{-1}$~\AA$^{-1}$)} & \textbf{(\%)} & \textbf{($^\circ$)} \\ 
\hline 
Full FoV & 2449903 & (15.90 $\pm$ 0.07) $\times$ 10$^{-16}$ & 4.3 $\pm$ 0.7 & 36.3 $\pm$ 5.0 \\
(6) & 2450295 & (14.67 $\pm$ 0.07) $\times$ 10$^{-16}$ & 1.4 $\pm$ 0.8 & 165.3 $\pm$ 14.7 \\
~ & 2450643 & (14.15 $\pm$ 0.07) $\times$ 10$^{-16}$ & 3.6 $\pm$ 0.8 & 2.5 $\pm$ 5.5 \\
~ & 2451014 & (12.97 $\pm$ 0.07) $\times$ 10$^{-16}$ & 2.1 $\pm$ 0.8 & 24.3 $\pm$ 10.9 \\
~ & 2451340 & (13.05 $\pm$ 0.06) $\times$ 10$^{-16}$ & 2.9 $\pm$ 0.8 & 25.1 $\pm$ 7.5 \\
\hline 
Nucleus & 2449903 & (56.91 $\pm$ 0.18) $\times$ 10$^{-17}$ & 4.4 $\pm$ 0.5 & 104.9 $\pm$ 3.2 \\
(0.49) & 2450295 & (50.67 $\pm$ 0.17) $\times$ 10$^{-17}$ & 7.9 $\pm$ 0.6 & 119.4 $\pm$ 1.9 \\
~ & 2450643 & (49.11 $\pm$ 0.17) $\times$ 10$^{-17}$ & 5.3 $\pm$ 0.6 & 126.1 $\pm$ 2.9 \\
~ & 2451014 & (43.11 $\pm$ 0.16) $\times$ 10$^{-17}$ & 6.5 $\pm$ 0.6 & 115.3 $\pm$ 2.5 \\
~ & 2451340 & (44.45 $\pm$ 0.16) $\times$ 10$^{-17}$ & 5.5 $\pm$ 0.6 & 119.6 $\pm$ 2.9 \\
\hline 
HST-1 & 2449903 & (47.18 $\pm$ 0.63) $\times$ 10$^{-18}$ & 22.8 $\pm$ 2.1 & 121.0 $\pm$ 2.7 \\
(0.25) & 2450295 & (39.95 $\pm$ 0.59) $\times$ 10$^{-18}$ & 17.5 $\pm$ 2.3 & 118.8 $\pm$ 3.9 \\
~ & 2450643 & (36.73 $\pm$ 0.57) $\times$ 10$^{-18}$ & 18.2 $\pm$ 2.4 & 124.7 $\pm$ 3.9 \\
~ & 2451014 & (38.02 $\pm$ 0.57) $\times$ 10$^{-18}$ & 12.1 $\pm$ 2.4 & 113.1 $\pm$ 5.6 \\
~ & 2451340 & (44.89 $\pm$ 0.57) $\times$ 10$^{-18}$ & 16.5 $\pm$ 2.0 & 122.5 $\pm$ 3.5 \\
\hline 
HST-2 & 2449903 & (56.68 $\pm$ 3.20) $\times$ 10$^{-19}$ & 6.5 $\pm$ 7.9 & 132.1 $\pm$ 39.8 \\
(0.15) & 2450295 & (46.32 $\pm$ 2.91) $\times$ 10$^{-19}$ & 7.9 $\pm$ 9.1 & 133.5 $\pm$ 37.5 \\
~ & 2450643 & (40.88 $\pm$ 2.80) $\times$ 10$^{-19}$ & 10.5 $\pm$ 10.3 & 151.4 $\pm$ 30.1 \\
~ & 2451014 & (50.22 $\pm$ 3.06) $\times$ 10$^{-19}$ & 6.9 $\pm$ 9.0 & 130.8 $\pm$ 41.2 \\
~ & 2451340 & (34.57 $\pm$ 2.55) $\times$ 10$^{-19}$ & 11.7 $\pm$ 11.9 & 165.9 $\pm$ 27.0 \\
\hline 
DE & 2449903 & (12.25 $\pm$ 0.09) $\times$ 10$^{-17}$ & 6.7 $\pm$ 1.3 & 4.0 $\pm$ 4.9 \\
(0.35) & 2450295 & (11.39 $\pm$ 0.09) $\times$ 10$^{-17}$ & 6.7 $\pm$ 1.3 & 2.6 $\pm$ 5.1 \\
~ & 2450643 & (11.31 $\pm$ 0.09) $\times$ 10$^{-17}$ & 9.4 $\pm$ 1.3 & 1.9 $\pm$ 3.7 \\
~ & 2451014 & (10.33 $\pm$ 0.09) $\times$ 10$^{-17}$ & 5.7 $\pm$ 1.4 & 174.5 $\pm$ 6.3 \\
~ & 2451340 & (11.21 $\pm$ 0.09) $\times$ 10$^{-17}$ & 7.0 $\pm$ 1.3 & 175.5 $\pm$ 4.7 \\
\hline 
DM & 2449903 & (36.88 $\pm$ 0.60) $\times$ 10$^{-18}$ & 34.6 $\pm$ 2.6 & 19.3 $\pm$ 2.1 \\
(0.25) & 2450295 & (30.09 $\pm$ 0.56) $\times$ 10$^{-18}$ & 32.0 $\pm$ 3.1 & 17.7 $\pm$ 2.6 \\
~ & 2450643 & (29.13 $\pm$ 0.53) $\times$ 10$^{-18}$ & 30.8 $\pm$ 3.0 & 18.5 $\pm$ 2.6 \\
~ & 2451014 & (26.14 $\pm$ 0.52) $\times$ 10$^{-18}$ & 32.9 $\pm$ 3.3 & 13.5 $\pm$ 2.6 \\
~ & 2451340 & (22.43 $\pm$ 0.48) $\times$ 10$^{-18}$ & 36.1 $\pm$ 3.5 & 15.6 $\pm$ 2.6 \\
\hline 
DW & 2449903 & (46.64 $\pm$ 0.76) $\times$ 10$^{-18}$ & 46.2 $\pm$ 2.8 & 167.8 $\pm$ 1.5 \\
(0.35) & 2450295 & (46.64 $\pm$ 0.74) $\times$ 10$^{-18}$ & 47.7 $\pm$ 2.7 & 167.9 $\pm$ 1.5 \\
~ & 2450643 & (46.22 $\pm$ 0.73) $\times$ 10$^{-18}$ & 47.8 $\pm$ 2.6 & 165.0 $\pm$ 1.5 \\
~ & 2451014 & (45.51 $\pm$ 0.72) $\times$ 10$^{-18}$ & 48.6 $\pm$ 2.6 & 163.5 $\pm$ 1.5 \\
~ & 2451340 & (43.14 $\pm$ 0.69) $\times$ 10$^{-18}$ & 47.2 $\pm$ 2.7 & 169.0 $\pm$ 1.5 \\
\hline 
E & 2449903 & (43.48 $\pm$ 0.87) $\times$ 10$^{-18}$ & 11.0 $\pm$ 3.3 & 9.6 $\pm$ 7.6 \\
(0.42) & 2450295 & (43.78 $\pm$ 0.84) $\times$ 10$^{-18}$ & 14.6 $\pm$ 3.2 & 174.5 $\pm$ 5.5 \\
~ & 2450643 & (42.02 $\pm$ 0.83) $\times$ 10$^{-18}$ & 23.1 $\pm$ 3.2 & 165.5 $\pm$ 3.7 \\
~ & 2451014 & (41.87 $\pm$ 0.80) $\times$ 10$^{-18}$ & 16.0 $\pm$ 3.0 & 160.9 $\pm$ 5.3 \\
~ & 2451340 & (37.82 $\pm$ 0.76) $\times$ 10$^{-18}$ & 17.1 $\pm$ 3.3 & 178.1 $\pm$ 4.9 \\
\hline 
\end{tabular}
\caption{Flux and polarization of the different knots and regions in POS~1 examined in Fig.~\ref{Fig:POS~1_frise} as a function of time.}   
\label{Tab:Knots}     
\end{table*}

\subsection{Scanning the jet}
\label{POS~1:Scan}

On the POS~1 figures, the position angle of the jet is almost constant along the relativistic beam. Its width is also constant, as highlighted by the bright knots and the faint emission detectable in the inter-knot regions. This width is approximately half an arcsecond and it is known to be thinner than in the radio band \citep{Sparks1994,Pasetto2021}. By scanning the source from the nucleus to the edge of the polarization map using an aperture radius that is consistent with the jet's width, it is possible to determine how the total flux, polarized flux, polarization degree and angle vary within the jet. From the information brought by $\Psi$, one can reconstruct the averaged magnetic filed topology of the jet, to be used in parsec-scale simulations. The polarized flux (that is the multiplication of the total flux with $P$) is another important quantity as, since we are measuring synchrotron emission, the polarized flux can be directly linked to an increase in isotropic (random) B-fields or an increase of an ordered B-field if the polarized flux decreases or increases, respectively. 

\begin{figure*}
\centering
\hbox{\hspace{2.3cm} \includegraphics[width=.77\textwidth]{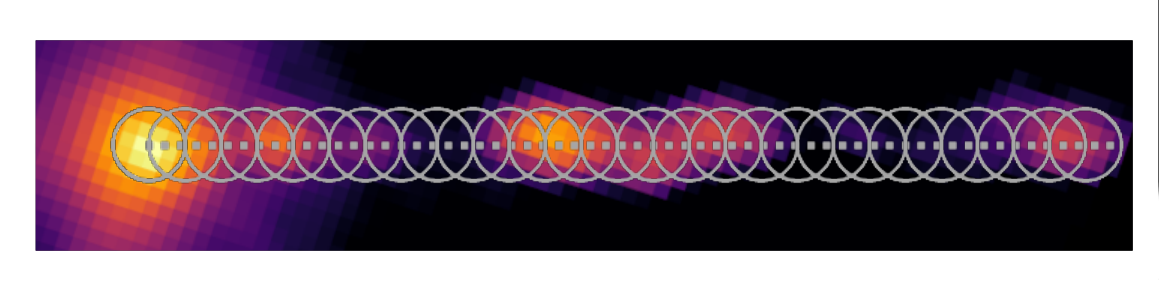}}
\includegraphics[width=.9\textwidth]{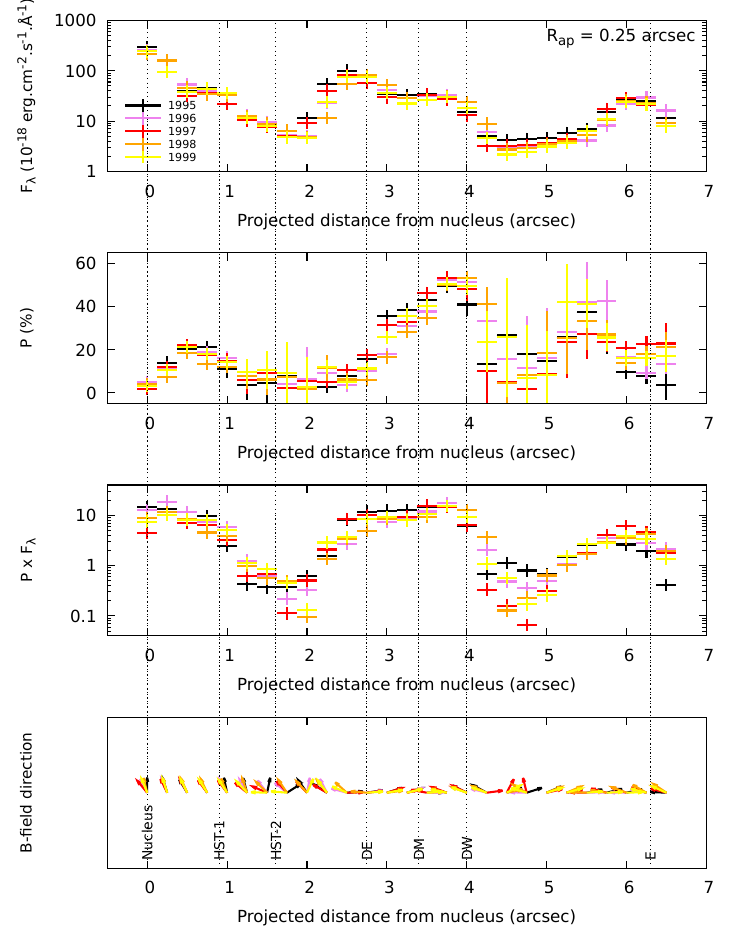}
\caption{Total flux $F$, polarization degree $P$, polarized flux ($P \times F$) and B-field direction (perpendicular to the polarization angle $\Psi$). The aperture radius used for the extraction zones is 0.25~arcseconds, covering the transversal width of the jet. The different epochs covered by the HST/FOC campaign are color-coded.}
\label{Fig:Across_the_jet}
\end{figure*}

We show in Fig.~\ref{Fig:Across_the_jet} (top) all the extraction regions across the jet, equally spaced by 0.25~arcsecond. The overlap of the extraction zones ensures a linearity in the properties probed by this scan. The total flux, polarization degree, polarized flux and B-field direction across the jet are presented in Fig.~\ref{Fig:Across_the_jet} (bottom). We repeated this scan for each of the five observation campaigns in order to detect significant variations within the jet, without specifically focusing on the knots. From the nucleus to the extremity of the jet shown in POS~1 : 

\begin{itemize}
    \item The observed total flux is decreasing with the projected distance from the core, except at the positions of the DE, DM and DW knots, and, further away, the E knot. The strongest flux variations between 1995 and 1999 occur not at the position of the bright features but rather in the inter-knot regions: right before DE (by factor 4) and E (by factor 2). 
    \item Interestingly, the polarization degree across the jet also shows strong variations, but they are not obligatory correlated to the position of a bright knot. $P$ increases from the nucleus to the HST-1 knot, up to 20\%, then decreases down to almost zero in the inter-knot region between HST-2 and DE. From knot DE, the polarization degree starts to increase up to a maximum of $\sim 50\%$ at the location of knot DW, then suddenly decreases down to zero again. One could expect, as for the inter-knot region between HST-2 and DE, that $P$ would stay almost null in the inter-knot region between DW and E. In contrary, $P$ rises as high as in the DW feature and then decreases at the location of knot E. Here, $P$ shows strong variation, from null to $\sim$ 20\%.
    \item The radial profile of the polarized flux is surprising as it shows smooth, wave-like variations across the jet. The peaks of polarized flux do not correlate with the peaks of total flux emission, nor with the peaks of $P$, but the lowest polarized fluxes are still associated with the inter-knot regions. Between knots HST-1 and HST-2, and at the location of knot E, the polarized flux increased between 1995 and 1999, possibly indicating an increase of unresolved anisotropic magnetic fields within those regions, due to some shocks or compression events in the jet. On the contrary, around knots DE and DM, and in the inter-knot region after DW, the polarized flux decreased between 1995 and 1999, which may indicate an increase of random (isotropic) magnetic fields, which may be turbulent.    
    \item Finally, the polarization position angle appears to be essentially parallel to the jet axis within the first 1.5 arcseconds from the nucleus and then mostly perpendicular to it. It means that the B-field is transverse close to the AGN core and then aligns with the jet propagation direction. The variation in $\Psi$ observed in the inter-knot regions (whre the polarized flux is at its lowest) are essentially due to the low count-rate statistics in those flux-deprived zones and should not be trusted.
\end{itemize}

We note an interesting similarity in polarization trend between our analysis and the one presented in \citet{Pasetto2021}. The authors published high resolution 4-18~GHz radio images of the M87 jet, with a particular focus on polarization. Although the data shown in \citet{Pasetto2021}'s paper mostly focus on POS~2, a region of the jet which was not observed by the FOC we recall, some knots are common in the two analyzes : knots D and E. Looking at Figs.~2 and 4 in \citet{Pasetto2021}, one can see that the radio polarization degree is of the order of 20-50\% in knot D, then decreases down to $\sim$ 20\% before reaching a value as high as 50-60\% in the inter-knot region. Further away from the nucleus, $P$ decreases again down to $\sim$ 35\% in knot E. To explain such variations, the authors suggested a double-helix morphology of the jet material : where the filaments are well separated the magnetic field is well ordered, reaching high fractional polarization values, while where the filaments intersect each other the emission suffers for strong depolarization effects. The radio polarization trend and values are similar to what was found in this section (see Fig.~\ref{Fig:Across_the_jet}), strengthening the idea that a scenario in which the jet is not a simple helix could be at work here. We will explore this hypothesis in the Discussion section.

\section{Analysis : POS~3}
\label{POS~3}

\subsection{Knot-by-knot examination}
\label{POS~3:clouds}

\begin{figure*}
\centering
\includegraphics[trim = 0.05cm 0.08cm 0.05cm 0.05cm, clip, width=.65\textwidth]{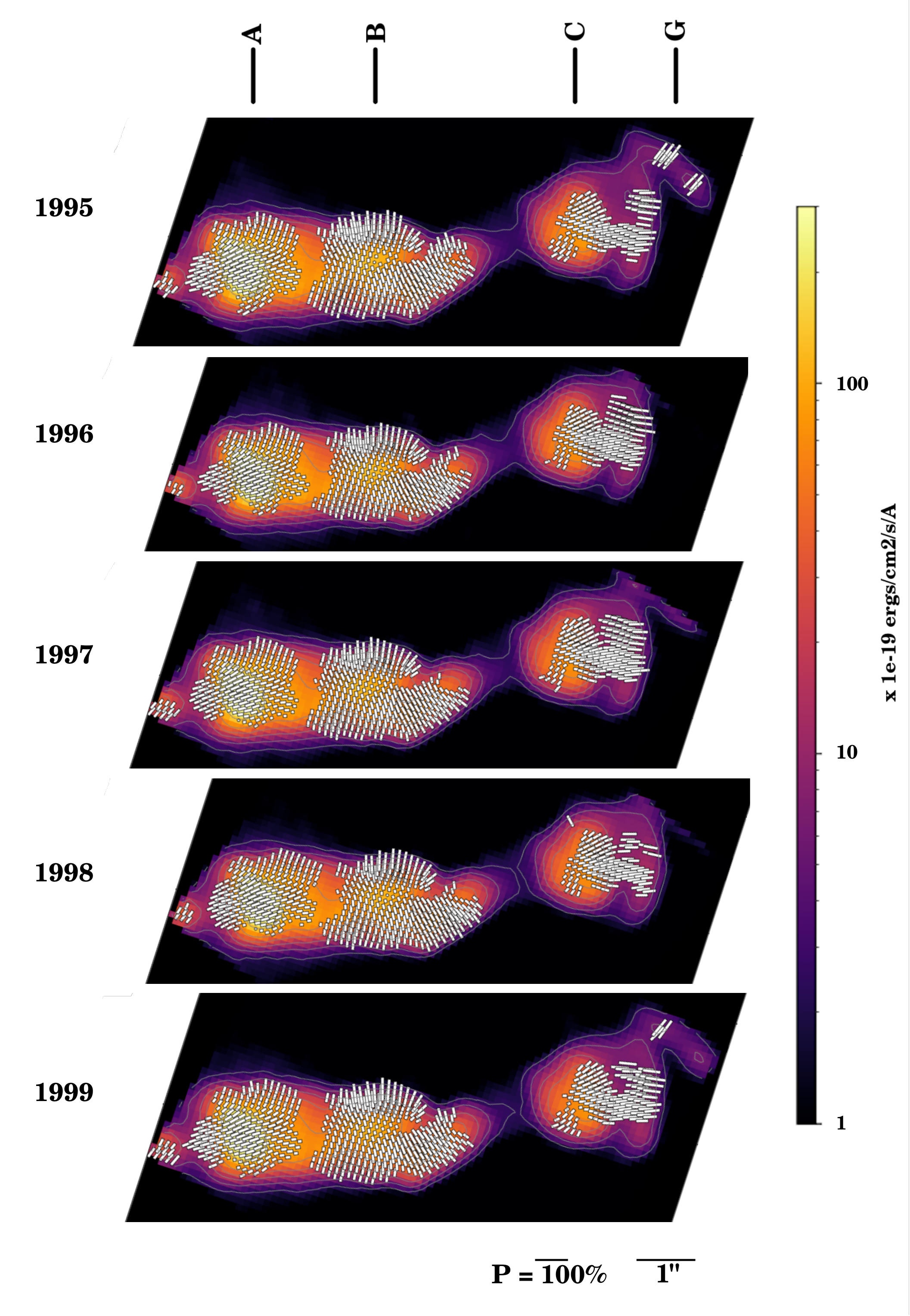}
\caption{Same as Fig.~\ref{Fig:POS~1_frise} but for POS~3.}
\label{Fig:POS~3_frise}
\end{figure*}

We now analyze the POS~3 polarization maps, presented in full scale in Appendix~\ref{AppendixB}. Four bright knots are identified, following the usual nomenclature. From the East to the West are knots A, B, C and G, presented in the chronological timeline of Fig.~\ref{Fig:POS~3_frise}. Knot A is the brightest in total flux, while G is the dimmest. Due to variation in pointing accuracy, knot G has been substantially cropped off the HST/FOC observation in 1996, 1997 and 1998. Tab.~\ref{Tab:Knots3} lists the flux, polarization degree and angle of each of the knots as a function of time. 

By focusing first on the full FoV, we see from Tab.~\ref{Tab:Knots3} that the total flux is slowly decreasing with time, as observed for POS~1, by a factor 1.15. On the other hand, both $P$ and $\Psi$ are constant in time, with subtle variations that could be associated to instrumental and pointings effects. On average, the POS~3 region is polarized at about 6\% with a polarization position angle of $\sim$ 144$^\circ$.

Knot A shows a flux that is slightly decreasing with time and accounts for a bit more than 40\% of the FoV flux. The polarization degree integrated in a 0.76~arcsec radius aperture is as high as 25\% with a polarization angle of $\sim$ 131$^\circ$. The magnetic structure, revealed by the orientation of the $\Psi$ vectors, is rather complex. The magnetic field lines are perpendicular to the jet, except on the surfaces of the knot, where it interacts with the intergalactic medium. The magnetic fields become parallel, highlighting the pressures forces acting at the boundaries of the knot. At such places, $P$ rises up to 45\% per spatial bin. Such rise is associated with a peak of polarized flux, revealing a compressed anisotropic magnetic field. 

Knot B, also showing a decreasing trend in total flux, shows a completely different polarization pattern. The magnetic field lines are now parallel to the jet propagation axis ($\Psi \sim 28^\circ$), while $P$ remains similar ($\sim 23\%$). One can see, on the POS~3 images, that the region between knot A and B is completely depolarized. This is due to the superposition of the two orthogonal polarizations, perfectly cancelling each other. The magnetic topology remains constant (parallel) in the whole knot, but shows increasing values of $P$ at the boundaries with the intergalactic medium too. The polarization reaches values up to $56\%$ (South of knot B) and $70\%$ (North of knot B), as already reported by \citet{Capetti1997}. 

Knot C follows knot B after an interclump region that is depolarized too. It is not surprising to observe a decreasing trend in flux too and a polarization position angle that has rotated by $90^\circ$ a second time (hence the depolarization in the interclump region). $P$ is of the order of $18\%$ at $125^\circ$. The polarization vectors vary smoothly across knot C, in a wave-like fashion, revealing oblique magnetic field lines. Such wiggle was already noted by \citet{Fraix1989}, although at a much lower spatial resolution.

Finally, knot D could have been observed only twice, at the beginning and end of the 5 years campaign. Since the knot is at the border of the detector frame, one shall be cautious about over-interpreting the results. The following should therefore be taken with caution. The flux seems to have slightly decreased while both $P$ and $\Psi$ remained constant within the reporter error bars. The magnetic field lines are found to be perpendicular to the jet directly, similarly to knot A and C. The high values of $P$ could indicate a well-ordered magnetic field 

For all four knots, the reported polarization degrees and angles are in good agreement with past measurements with similar aperture radii reported in \citet{Visvanathan1981}, \citet{Fraix1989} and \citet{Fresco2020}. Overall, the polarization in POS~3 is rather stable with time, in contrary to POS~1.

\subsection{A combined map of POS~3}
\label{POS~3:all_map}

Because the jet's width varies strongly across the radial jet direction (1.25 -- 2.25~arcseconds), because the jet shows a strong curvature between regions B and C, and because complex polarization profiles are observed within the knots (see Figs.~\ref{Fig:POS~3_maps_1995} to \ref{Fig:POS~3_maps_1999}), we could not repeat the exercise shown in Sect.~\ref{POS~1:Scan}, i.e. a scan of the jet. However, because POS~3 is remarkably stable in time in terms of polarization properties (at least at the spatial scale reported here and unlike POS~1), we can combine the five observations into a single one to increase the statistics by a factor $\sqrt{5}$. 

\begin{figure*}[htp]
\centering
\includegraphics[width=\textwidth]{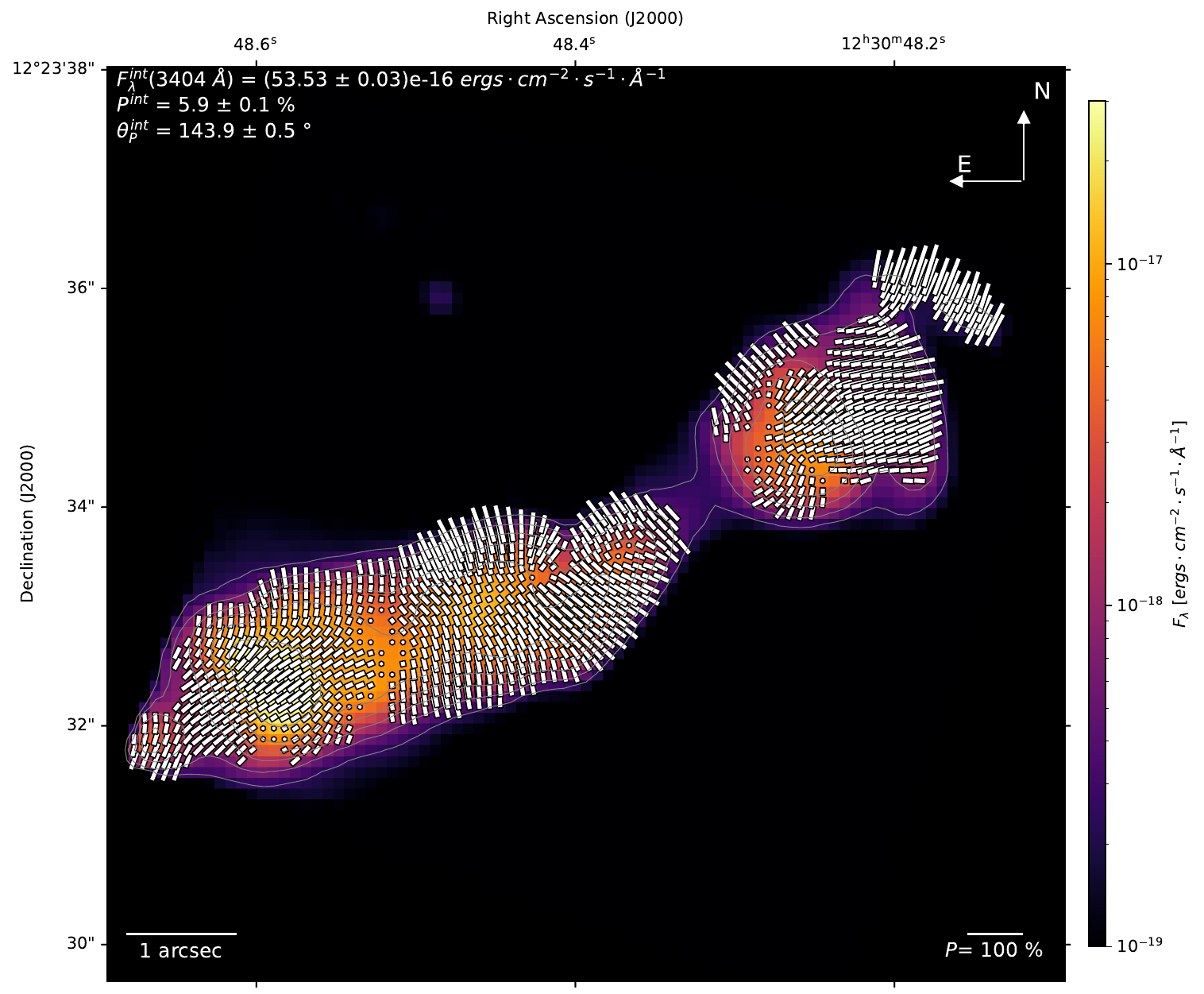}
\caption{Combined polarimetry map from the years 1995 to 1999 for POS~3. All technical details are similar to what was already written in the caption of Fig.~\ref{Fig:data_reduction}.}
\label{Fig:POS~3_all}
\end{figure*}

We show the resulting, combined map of POS~3 in Fig.~\ref{Fig:POS~3_all}. We used the same spatial binning and Gaussian smoothing as in all the POS~3 figures shown in appendices, the only difference being the five times longer integration time. We can notice that the number of bins with $\left[\text{S/N}\right]_P \geq 3$ has greatly increased, resulting in more white vectors plotted along the jet structure. Even more interestingly, we can follow the evolution of magnetic field lines much better thanks to the additional polarization vectors. At this spatial scale, the field lines form continuous patterns which are only broken at certain key points which correspond to where there is an orthogonal rotation of the polarization vectors (and of the magnetic fields) and therefore a complete depolarization of the medium. This is particularly visible between knots A and B, between knots B and C, but also within the knots themselves. In particular, inside the knot C, there is a zone located close to the inner northern wall of the jet which is entirely depolarized. The outer layer of the jet, compressed by the pressure difference between the interstellar medium and that of the jet, presents high $P$ and field lines that are parallel to the jet propagation axis, then there is this sinuous, depolarized valley, and then, within the jet core, the magnetic fields is found to be perpendicular to the jet propagation. At the depolarized location, there must undoubtedly be a strong anisotropy of the field lines. It should also be noted that there is no such depolarization valley in the southern part of knot C. Finally, the integrated flux and polarization values are in-line with the values reported in Tab.~\ref{Tab:Knots3} comforting us in the conclusion that POS~3 is only very slightly variable in total and polarized fluxes.

\begin{table*}
\centering
\begin{tabular}{l c c c c}
\hline\hline 
\textbf{Region} & \textbf{Date} & \textbf{Total flux} & \textbf{$P$} & \textbf{$\Psi$} \\
\textbf{(aperture in arcsec)} & \textbf{(MJD)} & \textbf{(ergs~cm$^{-2}$~s$^{-1}$~\AA$^{-1}$)} & \textbf{(\%)} & \textbf{($^\circ$)} \\ 
\hline
Full FoV & 2449907 & (57.49 $\pm$ 0.07) $\times$ 10$^{-16}$ & 5.8 $\pm$ 0.2 & 146.1 $\pm$ 1.0 \\
(6) & 2450300 & (55.33 $\pm$ 0.07) $\times$ 10$^{-16}$ & 5.7 $\pm$ 0.2 & 142.4 $\pm$ 1.0 \\
~ & 2450661 & (53.22 $\pm$ 0.06) $\times$ 10$^{-16}$ & 5.7 $\pm$ 0.2 & 144.8 $\pm$ 1.0 \\
~ & 2451025 & (51.64 $\pm$ 0.06) $\times$ 10$^{-16}$ & 6.5 $\pm$ 0.2 & 143.1 $\pm$ 0.9 \\
~ & 2451340 & (49.93 $\pm$ 0.06) $\times$ 10$^{-16}$ & 5.7 $\pm$ 0.2 & 142.8 $\pm$ 1.0 \\
\hline
A & 2449907 & (24.71 $\pm$ 0.04) $\times$ 10$^{-16}$ & 23.2 $\pm$ 0.3 & 131.8 $\pm$ 0.3 \\
(0.76) & 2450300 & (24.14 $\pm$ 0.04) $\times$ 10$^{-16}$ & 22.9 $\pm$ 0.3 & 133.2 $\pm$ 0.3 \\
~ & 2450661 & (22.38 $\pm$ 0.03) $\times$ 10$^{-16}$ & 22.8 $\pm$ 0.3 & 131.3 $\pm$ 0.4 \\
~ & 2451025 & (21.27 $\pm$ 0.03) $\times$ 10$^{-16}$ & 24.9 $\pm$ 0.3 & 130.4 $\pm$ 0.3 \\
~ & 2451340 & (20.98 $\pm$ 0.03) $\times$ 10$^{-16}$ & 23.2 $\pm$ 0.3 & 130.6 $\pm$ 0.4 \\
\hline
B & 2449907 & (15.52 $\pm$ 0.03) $\times$ 10$^{-16}$ & 24.4 $\pm$ 0.3 & 28.7 $\pm$ 0.4 \\
(1.0) & 2450300 & (14.90 $\pm$ 0.03) $\times$ 10$^{-16}$ & 22.3 $\pm$ 0.3 & 29.7 $\pm$ 0.4 \\
~ & 2450661 & (14.45 $\pm$ 0.03) $\times$ 10$^{-16}$ & 24.1 $\pm$ 0.3 & 27.6 $\pm$ 0.4 \\
~ & 2451025 & (13.75 $\pm$ 0.03) $\times$ 10$^{-16}$ & 23.0 $\pm$ 0.3 & 26.9 $\pm$ 0.4 \\
~ & 2451340 & (13.41 $\pm$ 0.03) $\times$ 10$^{-16}$ & 23.4 $\pm$ 0.3 & 27.7 $\pm$ 0.4 \\
\hline
C & 2449907 & (53.21 $\pm$ 0.16) $\times$ 10$^{-17}$ & 17.4 $\pm$ 0.5 & 125.4 $\pm$ 0.8 \\
(0.6) & 2450300 & (52.56 $\pm$ 0.16) $\times$ 10$^{-17}$ & 18.7 $\pm$ 0.5 & 122.6 $\pm$ 0.8 \\
~ & 2450661 & (50.67 $\pm$ 0.16) $\times$ 10$^{-17}$ & 19.0 $\pm$ 0.5 & 126.4 $\pm$ 0.8 \\
~ & 2451025 & (50.02 $\pm$ 0.16) $\times$ 10$^{-17}$ & 18.0 $\pm$ 0.5 & 127.5 $\pm$ 0.8 \\
~ & 2451340 & (46.52 $\pm$ 0.15) $\times$ 10$^{-17}$ & 17.7 $\pm$ 0.5 & 125.5 $\pm$ 0.9 \\
\hline
G & 2449907 & (23.68 $\pm$ 0.60) $\times$ 10$^{-18}$ & 59.4 $\pm$ 4.2 & 154.4 $\pm$ 2.0 \\
(0.46) & 2450300 & -- & -- & -- \\
~ & 2450661 & -- & -- & -- \\
~ & 2451025 & -- & -- & -- \\
~ & 2451340 & (22.32 $\pm$ 0.51) $\times$ 10$^{-18}$ & 57.0 $\pm$ 3.7 & 154.1 $\pm$ 1.8 \\
\hline
\end{tabular}
\caption{Flux and polarization of the different knots and regions in POS~3 examined in Fig.~\ref{Fig:POS~3_frise} as a function of time.}
\label{Tab:Knots3} 
\end{table*}

\section{Discussion}
\label{Discussion} 

\subsection{A brief summary of observations} 
\label{Discussion:summary}

The analysis of polarized maps obtained from POS~1 and POS~3 provides a comprehensive view of the temporal and spatial dynamics within M87's jet, shedding light on the intricate interplay between magnetic fields and jet morphology. 
Despite limited morphological changes over time, detailed scrutiny reveals significant fluctuations in polarization degrees and angles across various knots along the jet :
\begin{itemize}
  \item in POS~1, although minimal morphological alterations are observed over the observation period, pronounced variations in jet polarization degrees and angles are detected, especially in the nucleus and in the DE, DM, and DW knots, but also in the inter-knot region between DW and E knots. Interestingly, polarization is also detected near the AGN core, exhibiting stronger temporal variability than in most parts of the jet. 
  The absence of direct correlation between polarization properties and total flux suggests complex magnetic field dynamics decoupled from changes in jet luminosity;
  \item in POS~3, the overall morphology remains relatively stable in time for both the total and polarized fluxes. What is different is that we detected notable spatial changes in polarization properties within individual knots themselves, thanks to their larger sizes. Knot A exhibits a complex magnetic field topology, with perpendicular alignments near its boundaries indicative of interactions with the intergalactic medium. Knots B and C display distinct polarization patterns, with depolarization zones between and inside them, suggesting intricate magnetic field configurations. Knot D, although observed less frequently, also shows polarization characteristics consistent with the overall trends observed in other POS~3 knots.
\end{itemize}

These findings highlight the intricate nature of magnetic field structures within M87's jet, indicating localized (sub-arcsecond) variations in alignment and intensity. The observed decoupling between polarization properties and total flux underscores the importance of polarimetric observations in unraveling the underlying physics driving AGN jet magnetic topology and evolution.

\subsection{An heuristic model derived from the observations} 
\label{Discussion:model}

The emission and the polarization are distinguishable between the jet segment in POS1 and in POS3. One of the key features is that jet segments appear to be more collimated in POS1 (and in POS2, see \citealt{Sparks1996} and, on a larger spatial scale, \citealt{Pasetto2021}) than in POS3. It is difficult to reconcile all the observational differences in emission fluxes and polarization if we impose that POS3 is simply a continuation of jet flow from POS1 and POS2. It is also difficult to envision the continuous energizing of charged particles along the entire jet, yet the jet flow does not show a strong signature of being slowed down. To obtain a consistent picture that takes into account the HST observations presented here and other observations (e.g., \citealt{Sparks1996}, \citealt{Biretta1999}, \citealt{Perlman2011}, \citealt{Pasetto2021}), we sought an alternative to the paradigm invoking only simple shock(s) in the jet.

The thinner emission spine detected in the radio observations \citep[][]{Owen1989}, together with the HST observations of the different structural morphology in the POS1 and POS3 fields as presented in this work, strongly indicates that the jet could have co-axial structures, with different kinetic and dynamic properties. When the flow in the jet has a radial velocity gradient, or even a velocity discontinuity, with respect to the jet's symmetry axis, shear could develop within the jet along the flow direction \citep[see, e.g.,][]{Rieger2004}. While co-axial structures in the jet look appealing, a number of issues would need to be addressed properly before we may adopt it confidently to derive the explanation for the polarization properties of the M87 jet.

The simplest jet with a coaxial structure is that of the jet-in-jet scenario \citep[][]{Owen1989,Tavecchio2008}. An example theoretical realization of the jet-in-jet scenario is the magnetohydrodynamic (MHD) model described in \cite{Sobyanin2017MNRAS}. The inner jet, which is embedded in the outer jet, extracts energy from the rotating black hole \citep[][]{Blandford1977}, while the outer jet is powered by the accretion disk \citep[][]{Blandford1982}. The magnetic fields threading the inner and outer jets are expected to have helical structures. A recent observation \citep[][]{Pasetto2021} revealed that double helix configurations exist in the M87 jet. Helical magnetic fields have also been observed in jets in other AGN \citep[e.g., 3C273][]{Asada2002} and in X-ray binaries \citep[e.g., SS433][]{Roberts2008}. In the M87 jet, additional mechanisms may be present to aid the development of two distinguishable helices, each of which itself is an individual magnetic flux rope.

We further postulate that there is a strong density contrast between the inner jet and the outer jet. Also, the inner jet has higher speeds than the outer jet, which is partly a consequence of their differences in mass loading, and it has a low value of the plasma beta, i.e., $\beta \ll 1$ (thermal-to-magnetic energy ratio). This can be justified if the inner jet threads directly into the ergosphere and onto the black hole event horizon \citep[see, e.g.,][]{Thorne1986} and the outer jet is anchored to the magnetised flow in the accretion disk \citep[see, e.g.,][]{Punsly2001}. This implies that the Lorentz force dominates the thermal and plasma pressures, resulting in a force-free situation, and the magnetic field ${\boldsymbol B}$ in the inner jet satisfies the Helmholtz equation $(\nabla^2 + \mu^2)\;\!{\boldsymbol B} = 0$. In the linear force-free approximation, $\mu$ is uniform and is known as the force-free constant. A double helix magnetic field configuration as that derived from observational fits corresponds to winding magnetic flux ropes, which is one of the excited modes in the \citet{Freidberg2014} solution to the force-free equation \citep[see, e.g., discussions in][]{Hu2021}\footnote{Note that in the fit shown in \cite{Hu2021}, one of the helices is winding upward and the other one is winding downward. This is due to the different polarity of the two flux ropes. Here, the two helical flux ropes can both be winding upward (or downward), provided the current flows corresponding to the two flux ropes are in the same direction. In this case, it will require a return current outside the inner jet for current loop closure. Current closure is also required for other co-axial jet-in-jet models \citep[e.g.,][]{Sobyanin2017MNRAS,Gabuzda2018}.}
The double helix magnetic flux ropes preserve the helical symmetry in the jet, yet allow the $z$ translation symmetry along the jet (here $z$ is the direction designated to the local orientation of the jet).

The presence of the two dominant interlocking helical flux ropes with low plasma $\beta$ has several immediate consequences: \begin{itemize} 
    \item Global magnetic fields are maintained by currents. The double helix configuration of magnetic flux ropes implies a significant toroidal field component in the M87 jet, and hence a strong current flow \citep[cf. the coaxial current flows discussed in][]{Gabuzda2018} in the inner jet in a direction parallel to the local jet orientation. 
    \item As the magnetic flux ropes are interlocking, the magnetic tension within one would hold the other from breaking away, implying that it is harder for them to be separated compared to a bunch of parallel twisting field lines. We speculate that the interlocking helical magnetic flux ropes may contribute to maintaining the collimation of the inner jet over a large scale, though the pressure from the outer jet, which practically acts as a sheath, may also contribute to keeping the lateral expansion of the inner jet. 
    \item Field-line exchange magnetic reconnection can occur within each of the magnetic flux ropes, and slip-running reconnection is an example \citep[see][]{Aulanier2006,Jing2017}. As slip-running reconnection is not localized, it can occur in the magnetic flux ropes along the entire jet. Moreover, the particle acceleration in this kind of reconnection, which would align with the orientation of the flux ropes \citep[see, e.g.,][]{Zhang2021}, derives the energy from the magnetic fields, which in turn extract energy from the rotation of the black hole. Thus, it would affect the kinetic energy of the plasma flow in the inner jet, which is presumably Poynting flux-dominated (due to the low plasma $\beta$). 
\end{itemize}

In this model, unlike the inner jet, the outer jet is coupled with the accretion disk. They are accretion-driven relativistic outflows \citep[][]{Blandford1982}, and they have been modelled exhaustively by numerical MHD simulations \citep[e.g.,][]{McKinney2014MNRAS,Ryan2018,Mizuno2021,Yang2024}. The black hole, which is a fast-spinning accretor, may play a part in launching the jet, by providing angular momentum \citep[see, e.g.,][]{Punsly2001}, but its event horizon, which is crucial for the launch of the inner jet, is not essential here. Observations show that weakly magnetised accreting neutron stars, e.g., in the X-ray binary Cir X-1, which is a type~I X-ray burster \citep{Tennant1986}, are able to launch relativistic jets \citep[][]{Fender2004}. The outer jet is substantially mass-loaded, though it can still be magnetically dominated. The bulk flow velocity of the outer jet is slower \citep[][]{Tavecchio2008}, with $v_{\rm bulk} \lesssim c$, where $c$ is the speed of light. Despite this, much of the momentum can be carried by gas flow instead of the Poynting flux of the embedded magnetic field.

The interface between the inner jet and the outer jet would be turbulent, as the Kelvin-Helmholtz (KH) instabilities are common in shear flow with a large velocity gradient. This leads to the possibility of particle acceleration in the boundary layer when turbulence and also tangential shocks are the agents \citep[e.g.,][]{Stawarz2002,Kimura2018}. Particle acceleration in this scenario would derive the energy from the kinetic energy of the jet flow, which in turn will cause drag on both the inner jet and the outer jet. Moreover, the models implicitly assumes that the inner jet and the outer jet essentially belong to the same fluid body but with flow speed being much faster in the interior jet spine than in the exterior jet sheath. In the model that we propose in this work, the inner jet and the outer jet are two distinctive flows of different nature, while KH instabilities may be present if there is a large shear velocity gradient in the fluid in the interface regions. However, KH instability can be significant if the density contrast between the inner jet and the outer jet is sufficiently large and the magnetic tension in the flux rope is sufficiently strong. The field reconnection in the magnetic flux ropes can explain the narrow radio emission jet spine observed by \cite{Owen1989} as well as models invoking shear boundary-layer acceleration. The difference is that unknowns such as turbulence and the development of KH instabilities are options rather than necessities. The remaining question now is: "can the HST/FOC ultraviolet imaging polarimetric data be explained with the inner jet as a low plasma $\beta$ relativistic flow confined in interlocking double helix magnetic flux ropes"?

We start with the relative morphology of the flux images of jet segments in POS1 and POS3. First of all, the jet appears to be at least twice thinner in POS1 and thicker in POS3. Secondly, there is a strong brightening head, which has substantial thickness, in the jet segment in the POS3 images. We attribute the emission of the jet in POS1 images to emission from charged particles accelerated in slip-run magnetic reconnection in the inner jet. As this emission does not extract energy from the mass-load MHD flow, the outer jet retains its energy until its eventual encounter with a density barrier, where a strong shock is formed. The thick brightening head in POS~3 is emission from the shock-accelerated charged particles.

We next look at the polarization properties along the jet in POS1. Traditionally, bright knots in AGN jets are attributed to multiple shocks. Generating multiple (oblique) shocks in the jet would require very specific settings \citep[see e.g.][]{Saxton2010MNRAS,Meli2013, Mandall2022ApJ}. Alternatively, they are attributed to the presence of "colliding shells" \cite[see e.g.][]{Mimica2007}. These two kinds of situations do not generally arise naturally within a large-scale highly collimated flow. Moreover, shock models have difficulties reconciling that the polarization vectors in the jet segment in POS1 do not indicate any signs that the magnetic fields are compressed such that they are perpendicular to the jet flows. Acceleration by shear boundary (between the inner and outer jet) shocks or turbulence \citep[e.g.][]{Stawarz2002} can accommodate the alignment of magnetic field along the jet segments, yet they have difficulties explaining why polarization in the dark regions between the bright knots could have stronger polarization degrees (see Tab.~\ref{Tab:Knots}). 

In the magnetic reconnection in interlocking "double (or even multiple) helix magnetic flux ropes" scenario, the magnetic field vectors inferred from the observed polarization would preferentially align with the jet segment orientation. The acceleration can occur globally along the jet. As twisting flux ropes are prone to kink instabilities \citep[see][]{Lyubarskii1999,Tchekhovskoy2016}, the bright knots would correspond to the kinks. The brightening can be due to a number of factors, such as the optical depth effect and enhancement of the local magnetic field. It can also be due to the fact that part of the magnetic flux rope is oriented in a direction more closely aligned to the line-of-sight, thus radiation from emitters streaming along the flux rope would be Doppler boosted. Furthermore, kinks can lead to magnetic reconnection\footnote{In \cite{Lazarian2019}, turbulence developed in magnetic field kinks leads to particle acceleration in GRB (gamma-ray burst) jet. In this scenario, the flux ropes are not required, though a helical magnetic field configuration in the jet is essential.}, and the subsequent enhancement in particle acceleration \citep[see][]{Ripperda2017,Davelaar2020ApJ} leads to local emission brightening. Because the toroidal magnetic field component is always present in addition to the magnetic field component parallel to the jet, the observed degree of polarization, if from an electron synchrotron process, from a segment of a flux rope cannot reach levels as high as 70\%, because polarimetry observations sample the emission associated with the parallel field component and the toroidal field component. In between the kinks, the magnetic flux ropes are stretched and restored by the magnetic tension force to align with the jet again, provided the two helical flux ropes do not develop kinks of the same scale/strength at the same location at the same time. Reconnection still occurs continually in the magnetic flux ropes. Though the emission there is not as strong as that in the kinks, the flux ropes are now aligned with the jet, and hence the observed degree of polarization is expected to exceed that of the emission from the kinks.

To explain the polarization properties of the emission from the jet segments in POS3, it would require the presence of a strong shock and reconnection in the draped magnetic fields in the post-shock flow. The formation of a strong shock when the outer jet encounters a density barrier is a natural consequence, given that the outer jet, which has not suffered significant loss in energy and momentum when it propagated, is expected to be relativistic and supersonic. At the shock front, the observed magnetic field would have a strong component perpendicular to the shock, consistent with the observed orientation of the polarization vectors in the brightest lump in POS3. Although the downstream flow may become turbulent, the magnetic configuration of interlocking flux ropes of the inner jet is more resilient, and there would be a magnetic drape enveloping the surviving inner jet. While the turbulence in the post-shock flow can accelerate particles, micro-scale magnetic reconnection (not necessarily the slip-running reconnection) can be triggered, which in turn can also accelerate particles \citep[see e.g.][for particle acceleration during reconnections in low $\beta$ plasma]{Li2017}. When energetic particles diffuse into the magnetic drapes (cf. magnetic field draping around the bow shock, magnetic sheath, and magnetopause configuration of the Earth after encountering the Solar wind), the field orientation is more ordered and roughly aligned with the jet flow, as indicated by the observed polarization vectors in fainter knots behind the very bright spot in POS~3.

We note that the presence of a slow, mass-load outer jet sheathing the inner jet would enforce collimation and suppress kinks
rather than contributing to the de-collimation of the faster inner jet. Provided that $(\rho v^2)|_{\rm inner-jet} > (\rho v^2)|_{\rm outer-jet}$, the outer jet will exert a lateral pressure onto the inner jet (Bernoulli's principle). Even when $(\rho v^2)|_{\rm inner_jet} > (\rho v^2)|_{\rm out_jet}$ is violated, the inner jet cannot expand and cause the outer jet to widen, because the tension force within the flux rope will keep the interlocking flux ropes in place. The interlocking between the flux ropes can only be destroyed when a large-scale wholesale reconnection is triggered by some violent processes, which provide the required strong current flows. We also note that helical jets are common \citep[][]{Gabuzda2014}. Helical jets, which are initially well-collimated with polarization vectors perpendicular to jet orientation, end with a strong terminated shock that has also been observed in stellar black hole systems, and an example is the microquasar SS433 \citep[see][]{Kaaret2024,Safi-Harb2022}. Thus, there is correspondence of jet physics in microquasars and in quasars (though M87 would need to be placed at a high redshift).

\subsection{Other sources in the fields-of view}
\label{Discussion:extra}

\begin{figure*}
\centering
\includegraphics[width=\columnwidth]{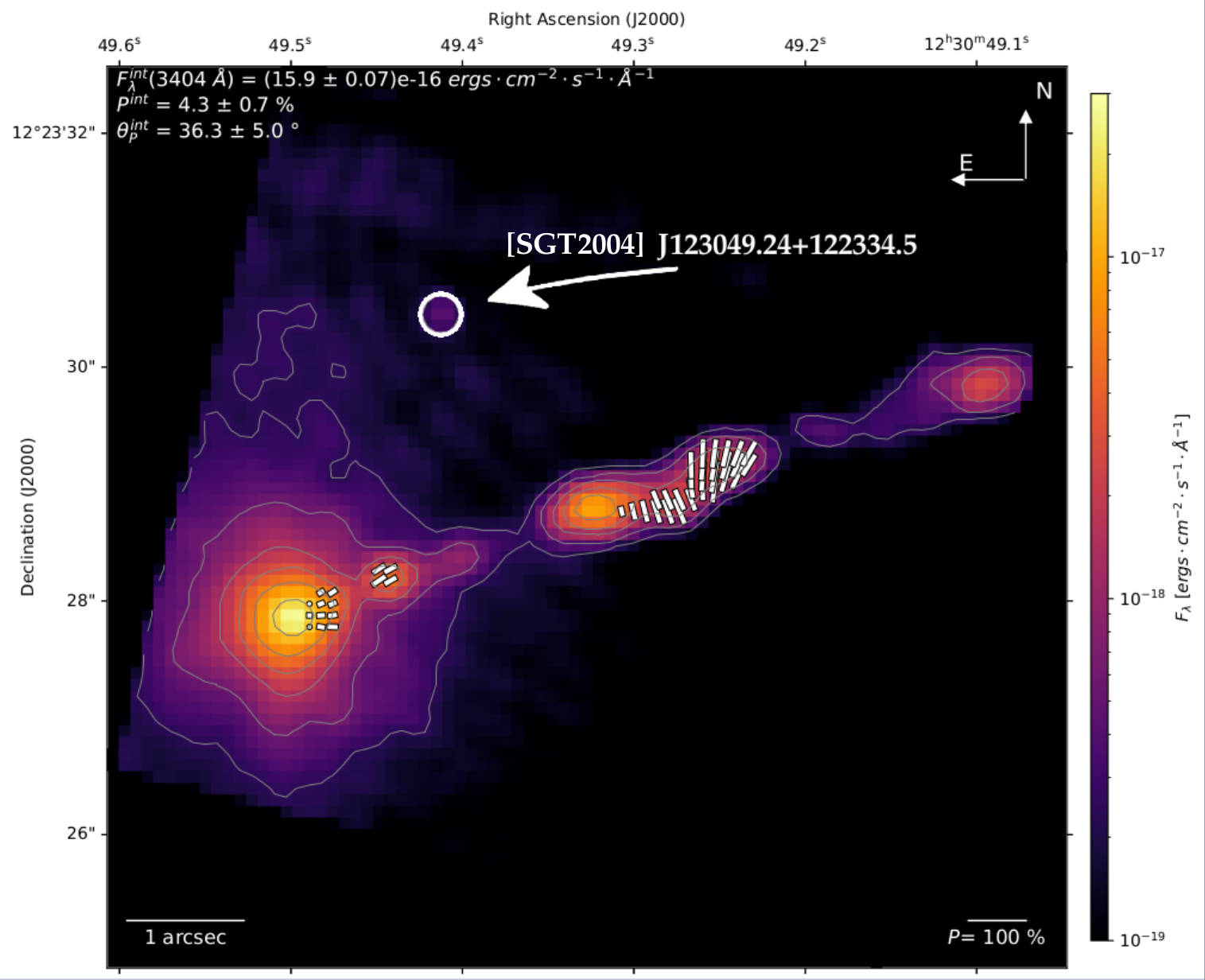}
\includegraphics[width=\columnwidth]{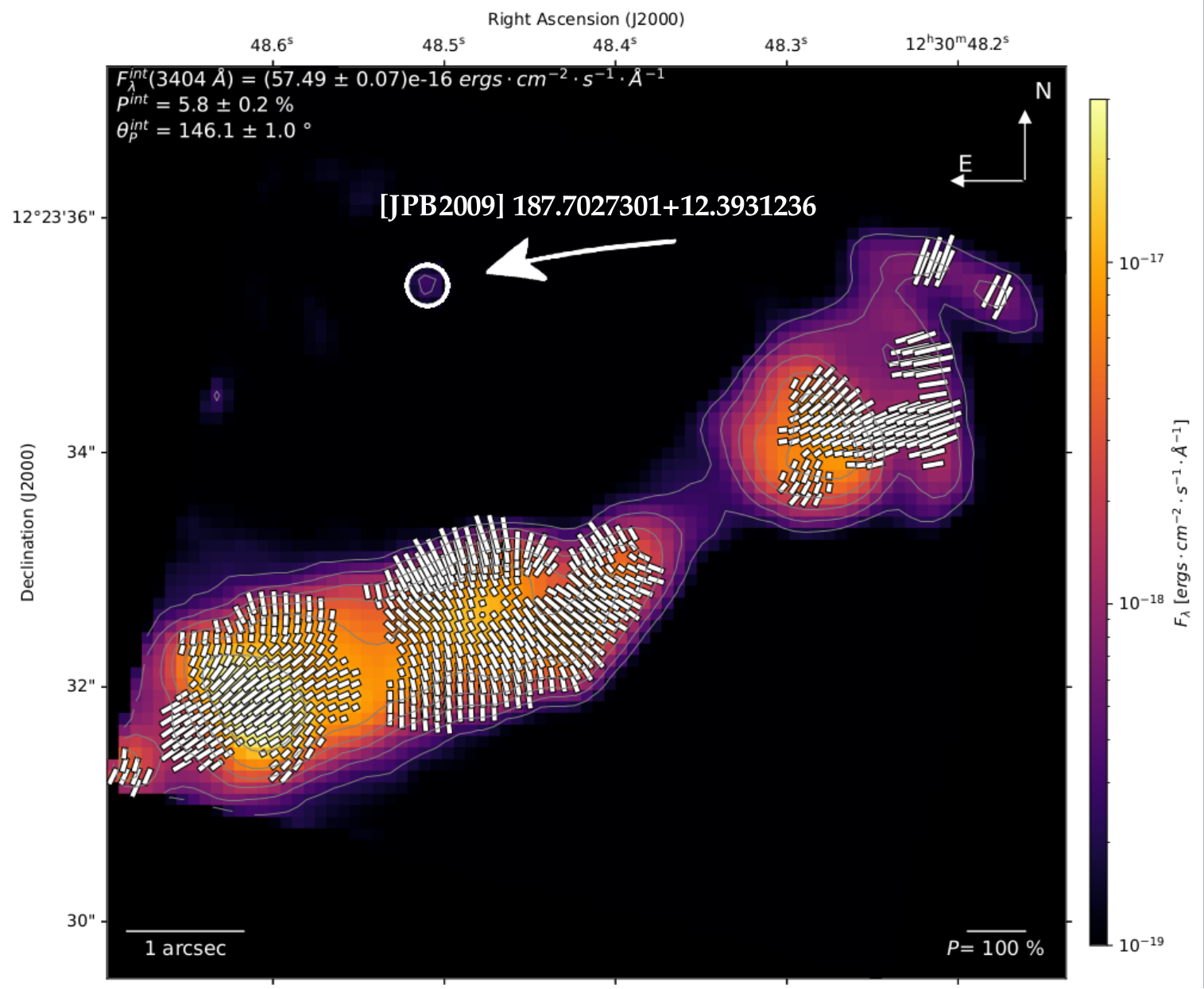}
\caption{HST/FOC 1995 observation of M87's POS~1 (left) and POS~3 (right) showing the extra sources in the fields of view. These sources, circled in white and indicated with a white arrow, are [SGT2004] J123049.24+122334.5 (left) and [JPB2009] 187.7027301+12.3931236 (right), see text for details.}
\label{Fig:POS_extra}
\end{figure*}

On the images of POS~1 and POS~3 appear two point sources which are completely uncorrelated to the M87 jet (see Fig.~\ref{Fig:POS_extra}). After scanning the FoV using the interactive sky atlas Aladin\footnote{\url{https://aladin.cds.unistra.fr/aladin-f.gml}}, we found that they correspond to :
\begin{itemize}
    \item (POS~1) an Ultraluminous X-ray source (ULX) called [SGT2004] J123049.24+122334.5 (RA = 187.7051666667$^\circ$, DEC = +12.3929166667$^\circ$) that is listed in the catalogue of \citet{Swartz2004}. It is one of the 154 discrete non-nuclear ULXs observed with the Chandra's Advanced CCD Imaging Spectrometer. 
    \item (POS~3) a Globular Cluster (GC) candidate called [JPB2009] 187.7027301+12.3931236 (RA = 187.702730100$^\circ$, DEC = +12.3931236000$^\circ$) that is listed in the catalogue of \citet{Jordan2009}. It is one of the globular cluster candidates detected among the 100 galaxies of the Advanced Camera for Surveys Virgo Cluster Survey. 
\end{itemize}

For the sake of curiosity and completeness, we simulated a 0.22~arcsecond aperture radius centered around both the ULX and the GC candidate and measured their UV flux and polarization for each of the five observation. We also summed (in Stokes parameters) the five observations per object to increase the statistics, despite potential intrinsic variability of the source polarization. This is equivalent to more than 6.5 hours of cumulative observation time with a space instrument equipped with a 2.4 meter diameter mirror. Tab.~\ref{Tab:ULX} reports the various measurements we obtained for the ULX and Tab.~\ref{Tab:GC} does the same for the GC candidate.

Starting with the ULX, we find that the source is quite faint, with an averaged flux of (40.93 $\pm$ 1.78) $\times$ 10$^{-19}$ erg cm$^{-2}$ s$^{-1}$ \AA$^{-1}$ at 3404~\AA. No polarization is measured during the first four years of observation and only upper limits could be set on $P$ but, in 1999, a poor detection is achieved (22.2\% $\pm$ 13.3\%). Summing the five years of observation leads to an upper limit of 10.9\%. It is merely impossible to compare these numbers to other ULX polarization because the polarization of ULXs is basically unknown whatever the wavelength band, with the only exception of the Galactic ULX pulsar Swift~J0243.6+6124 that was recently observed with the Imaging X-ray Polarimetry Explorer (IXPE) and the RoboPol polarimeter located in the focal plane of the 1.3 m telescope of the Skinakas observatory (Greece). \citet{Poutanen2024} and \citet{Majumder2024} have found that Swift~J0243.6+6124 X-ray polarization strongly varies with the pulsar phase and flux of the ULX, spanning from 2\% to 20\% at various polarization position angles. The optical polarization (R-band) is found to be very different : 1 -- 2.5\%, with a polarization angle of 20 -- 50$^\circ$, depending on the choice of the field star \citep{Poutanen2024}. No polarimetric measurements of an ULX have ever been achieved in the UV and our result, although mediocre due to the weakness of the source and the lack of statistics, is compatible with the aforementioned measurement for Swift~J0243.6+6124 and encourages new, more in-depth observations.

Regarding the GC candidate, its UV flux appears even fainter than that of the ULX (averaged value of (23.47 $\pm$ 1.14) $\times$ 10$^{-19}$ erg cm$^{-2}$ s$^{-1}$ \AA$^{-1}$ at 3404~\AA). No polarization was detected among the five years, with upper limits on $P$ as high as 53.7\% in 1999, when the flux of the source was the dimmest. The summed polarization is an upper limit of 9.0\%. Globular clusters, which are dense collections of old stars, are generally not known for having intrinsic polarization \citep{Martin1981,Hodge1996}. However, in some rare cases, polarization can be detected from GC due to interstellar polarization, where dust grains in the interstellar medium align with the Galactic magnetic field and polarize the light. This effect is not intrinsic to the clusters but results from the light's journey through space \citep{Minniti1990,Clayton2004}. Additionally, individual stars within globular clusters, particularly variable stars or evolved stars with circumstellar envelopes, may exhibit some polarization, though this is usually minimal when averaged across the cluster \citep{Shakhovskoi1964,Shakhovskoi1965}. Intra-cluster dust scattering could also cause polarization, but this is rare and more relevant to younger star clusters \citep{Martin1981}.

\begin{table}
\centering
\begin{tabular}{l c c c}
\hline\hline 
\textbf{Date} & \textbf{Total flux} & \textbf{$P$} & \textbf{$\Psi$} \\
\textbf{(year)} & \textbf{(ergs~cm$^{-2}$~s$^{-1}$~\AA$^{-1}$)} & \textbf{(\%)} & \textbf{($^\circ$)} \\ 
\hline
1995 & (36.7 $\pm$ 4.1) $\times$ 10$^{-19}$ & $\le$ 34.7 & -- \\
1996 & (42.6 $\pm$ 3.9) $\times$ 10$^{-19}$ & $\le$ 20.2 & -- \\
1997 & (38.3 $\pm$ 3.9) $\times$ 10$^{-19}$ & $\le$ 28.3 & -- \\
1998 & (37.5 $\pm$ 3.8) $\times$ 10$^{-19}$ & $\le$ 19.6 & -- \\
1999 & (38.9 $\pm$ 3.6) $\times$ 10$^{-19}$ & 22.2 $\pm$ 13.3 & 43.0 $\pm$ 19.4 \\
\hline
Combined & (40.93 $\pm$ 1.78) $\times$ 10$^{-19}$ & $\le$ 10.9 & -- \\
\hline
\end{tabular}
\caption{Flux and polarization of [SGT2004] J123049.24+122334.5, the point-source serendipitously appearing in the FoV of POS~1 (see Fig.~\ref{Fig:POS_extra}).}
\label{Tab:ULX} 
\end{table}

\begin{table}
\centering
\begin{tabular}{l c c c}
\hline\hline 
\textbf{Date} & \textbf{Total flux} & \textbf{$P$} & \textbf{$\Psi$} \\
\textbf{(year)} & \textbf{(ergs~cm$^{-2}$~s$^{-1}$~\AA$^{-1}$)} & \textbf{(\%)} & \textbf{($^\circ$)} \\ 
\hline
1995 & (26.2 $\pm$ 2.7) $\times$ 10$^{-19}$ & $\le$ 19.5 & -- \\
1996 & (27.2 $\pm$ 2.8) $\times$ 10$^{-19}$ & $\le$ 32.1 & -- \\
1997 & (23.1 $\pm$ 2.4) $\times$ 10$^{-19}$ & $\le$ 21.2 & -- \\
1998 & (24.3 $\pm$ 2.5) $\times$ 10$^{-19}$ & $\le$ 29.5 & -- \\
1999 & (16.7 $\pm$ 2.3) $\times$ 10$^{-19}$ & $\le$ 53.7 & -- \\
\hline
Combined & (23.47 $\pm$ 1.14) $\times$ 10$^{-19}$ & $\le$ 9.0 & -- \\
\hline
\end{tabular}
\caption{Flux and polarization of [JPB2009] 187.7027301+12.3931236, the point-source serendipitously appearing in the FoV of POS~3 (see Fig.~\ref{Fig:POS_extra}).}
\label{Tab:GC} 
\end{table}

\section{Conclusions}
\label{Conclusion} 

The analysis of the polarized maps of M87's jet from POS~1 and POS~3 revealed significant spatial and temporal fluctuations in polarization, suggesting complex magnetic field dynamics. The observed differences in collimation and polarization between POS~1 and POS~3 indicate that the jet may have co-axial structures with distinct kinetic properties. A model involving interlocking double helix magnetic flux ropes within a jet-in-jet structure nicely explains these observations. This model accounts for the magnetic field configurations, polarization properties, and overall jet morphology, highlighting the importance of polarimetric studies in understanding AGN jet dynamics. Alongside the analysis of the jet, we also report the measurement of the polarization of [SGT2004] J123049.24+122334.5 (an ULX) and [JPB2009] 739
187.7027301+12.3931236 (a GC candidate), for which we were able to place upper limits.

High-resolution imaging polarimetry, particularly in ultraviolet and radio wavelengths, is crucial for probing astrophysical jets. These techniques provide detailed information on the magnetic field structures and particle acceleration processes within jets. Ultraviolet polarimetry, in particular, not only offers insights into the high-energy processes along the jets but also within the immediate environments of supermassive black holes. Both wavebands can trace magnetic field configurations over larger scales, but at different timescales  : radio synchrotron photons being emitted latter than ultraviolet photons (the difference in emission timing being due to the energy loss mechanisms affecting the relativistic electrons responsible for synchrotron radiation). Together, these observations allow for a comprehensive understanding of the intricate physics governing astrophysical jets, making them indispensable tools in modern research. 

Unfortunately, high angular resolution ultraviolet polarization imaging is no longer available today and this paper therefore advocates the design of a new spatial instrument dedicated to this type of observational technique.

\vspace*{0.5cm}

\textbf{Data availability.} Tables 2-5 are available in electronic form at the CDS via anonymous ftp to cdsarc.u-strasbg.fr (130.79.128.5) or via http://cdsweb.u-strasbg.fr/cgi-bin/qcat?J/A+A/.

\vspace*{0.5cm}

\begin{acknowledgements}
The authors would like to acknowledge the anonymous referee for her/his comments that substancially improved the quality of this paper. F.M. and T.B. would like to acknowledge the support of the CNRS, the University of Strasbourg, the PNHE and PNCG. This work was supported by the "Programme National des Hautes Énergies" (PNHE) and the "Programme National de Cosmologie et Galaxies (PNCG)" of CNRS/INSU co-funded by CNRS/IN2P3, CNRS/INP, CEA and CNES. K.W. acknowledges the support from the UCL Cosmoparticle Initiative. E.L.-R. is supported by the NASA/DLR Stratospheric Observatory for Infrared Astronomy (SOFIA) under the 08\_0012 Program. SOFIA is jointly operated by the Universities Space Research Association,Inc.(USRA), under NASA contract NNA17BF53C, and the Deutsches SOFIA Institut (DSI) under DLR contract 50OK0901 to the University of Stuttgart. E.L.-R. is supported by the NASA Astrophysics Decadal Survey Precursor Science (ADSPS) Program (NNH22ZDA001N-ADSPS) with ID 22-ADSPS22-0009 and agreement number 80NSSC23K1585. This work has made use of the NASA Astrophysics Data System. 
\end{acknowledgements}

\bibliographystyle{aa} 
\bibliography{biblio} 

\begin{appendix}

\section{polarized maps of POS~1}
\label{AppendixA}

\begin{figure*}[htp]
\centering
\includegraphics[width=\textwidth]{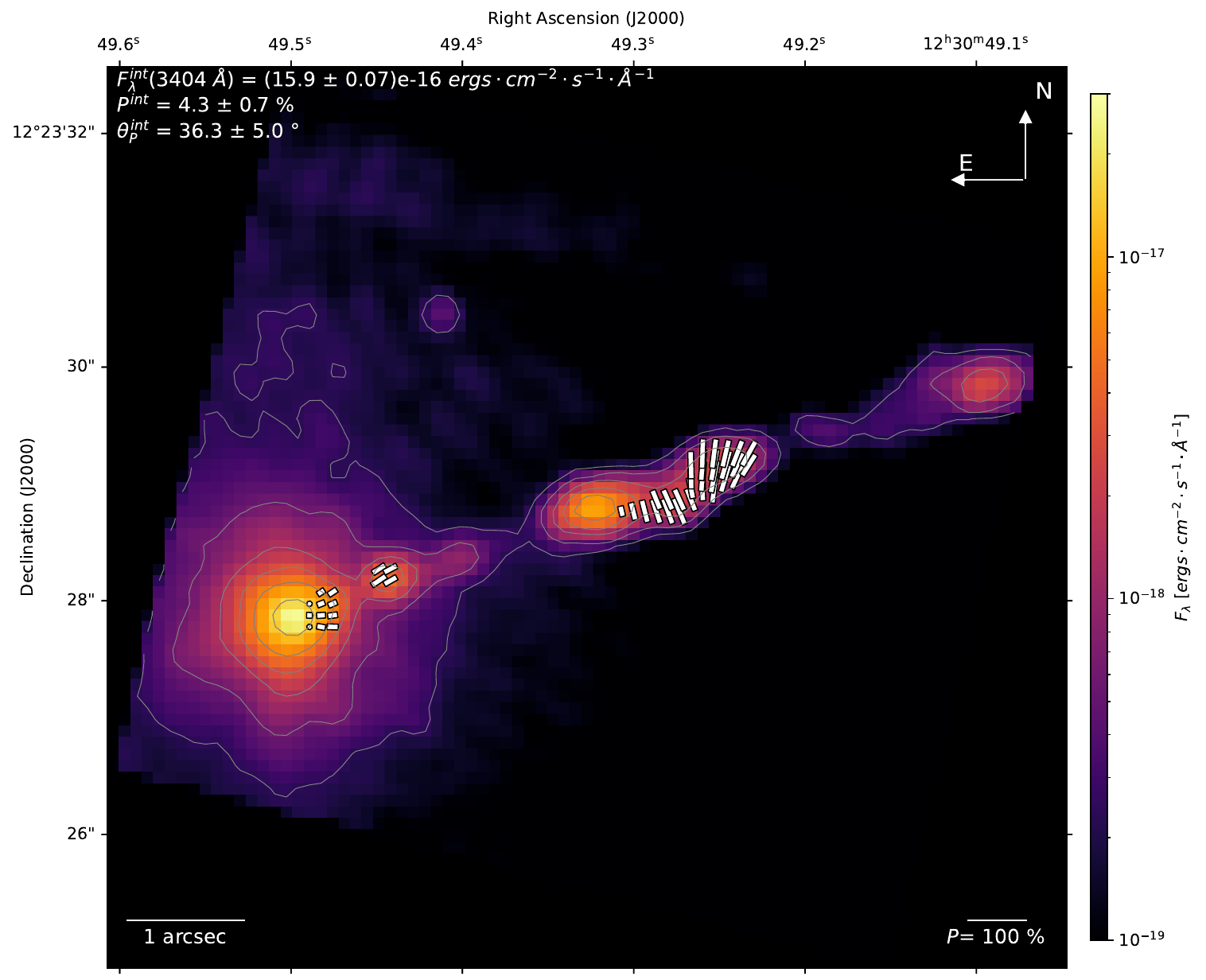}
\caption{HST/FOC 1995 observation of M87's POS~1 resampled to pixels of $0.1'' \times 0.1''$. Intensity is colour-coded in erg cm$^{-2}$ s$^{-1}$ \AA$^{-1}$. polarization vectors are displayed for $\left[\text{S/N}\right]_P \geq 3$. The contours are displayed for 0.8\%, 2\%, 5\%, 10\%, 20\% and 50\% of the maximum flux. On the top-left corner of the maps are displayed the flux and polarization values, integrated over the whole FOC FoV ($7'' \times 7''$).}
\label{Fig:POS~1_maps_1995}
\end{figure*}

\begin{figure*}[htp]
\centering
\includegraphics[width=\textwidth]{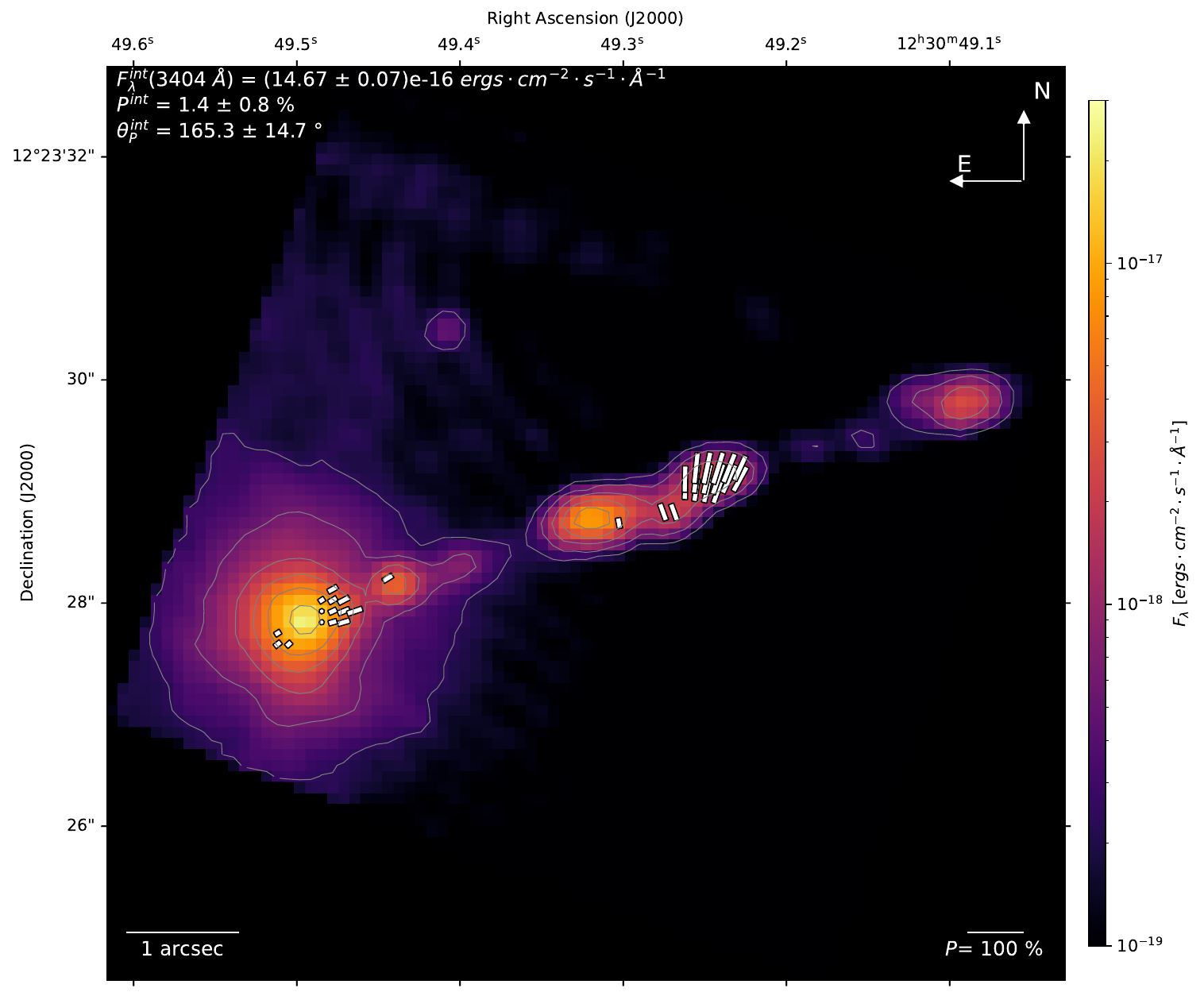}
\caption{Same as Fig.~\ref{Fig:POS~1_maps_1995}, but for the second observation (1996).}
\label{Fig:POS~1_maps_1996}
\end{figure*}

\begin{figure*}[htp]
\centering
\includegraphics[width=\textwidth]{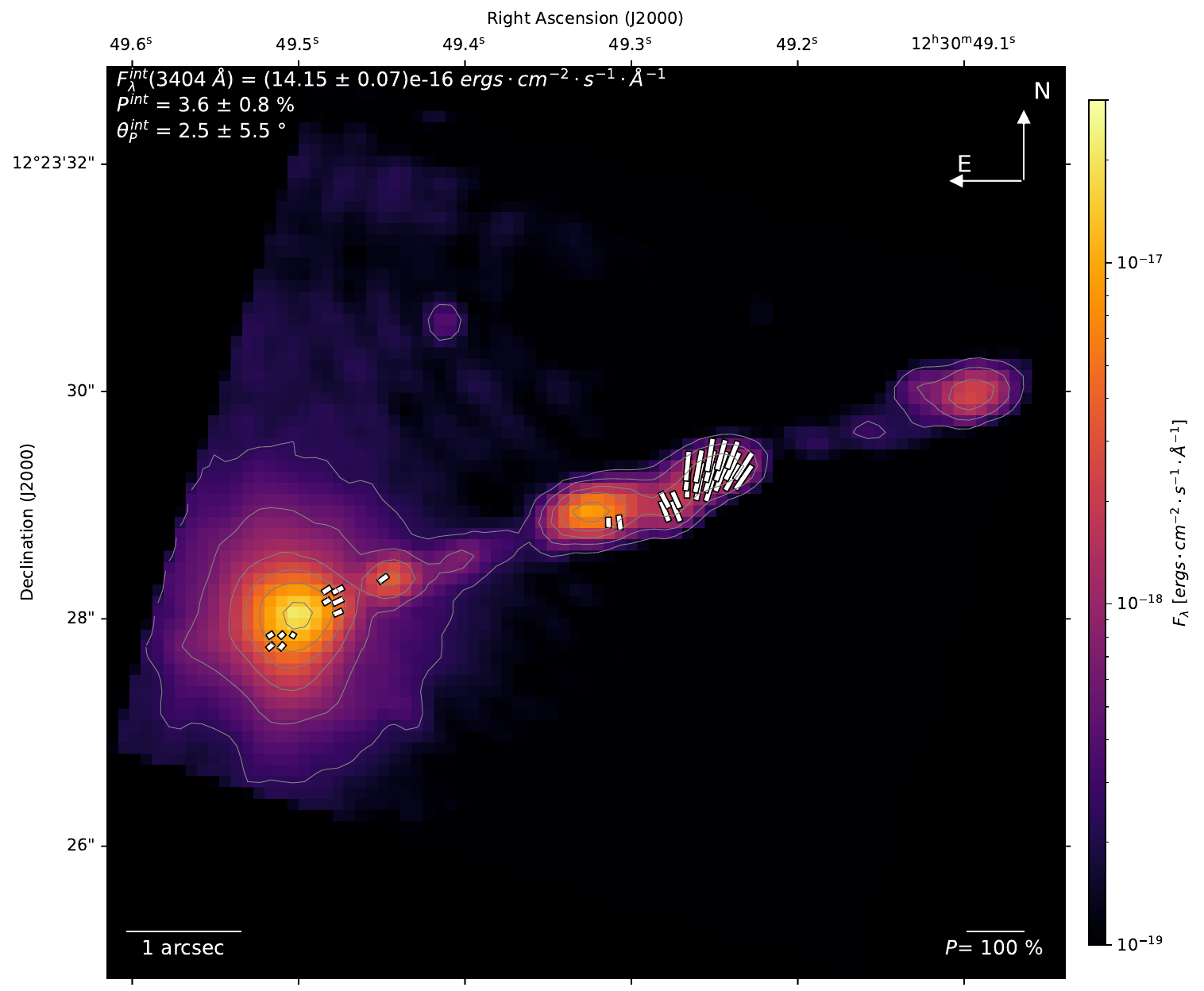}
\caption{Same as Fig.~\ref{Fig:POS~1_maps_1995}, but for the third observation (1997).}
\label{Fig:POS~1_maps_1997}
\end{figure*}

\begin{figure*}[htp]
\centering
\includegraphics[width=\textwidth]{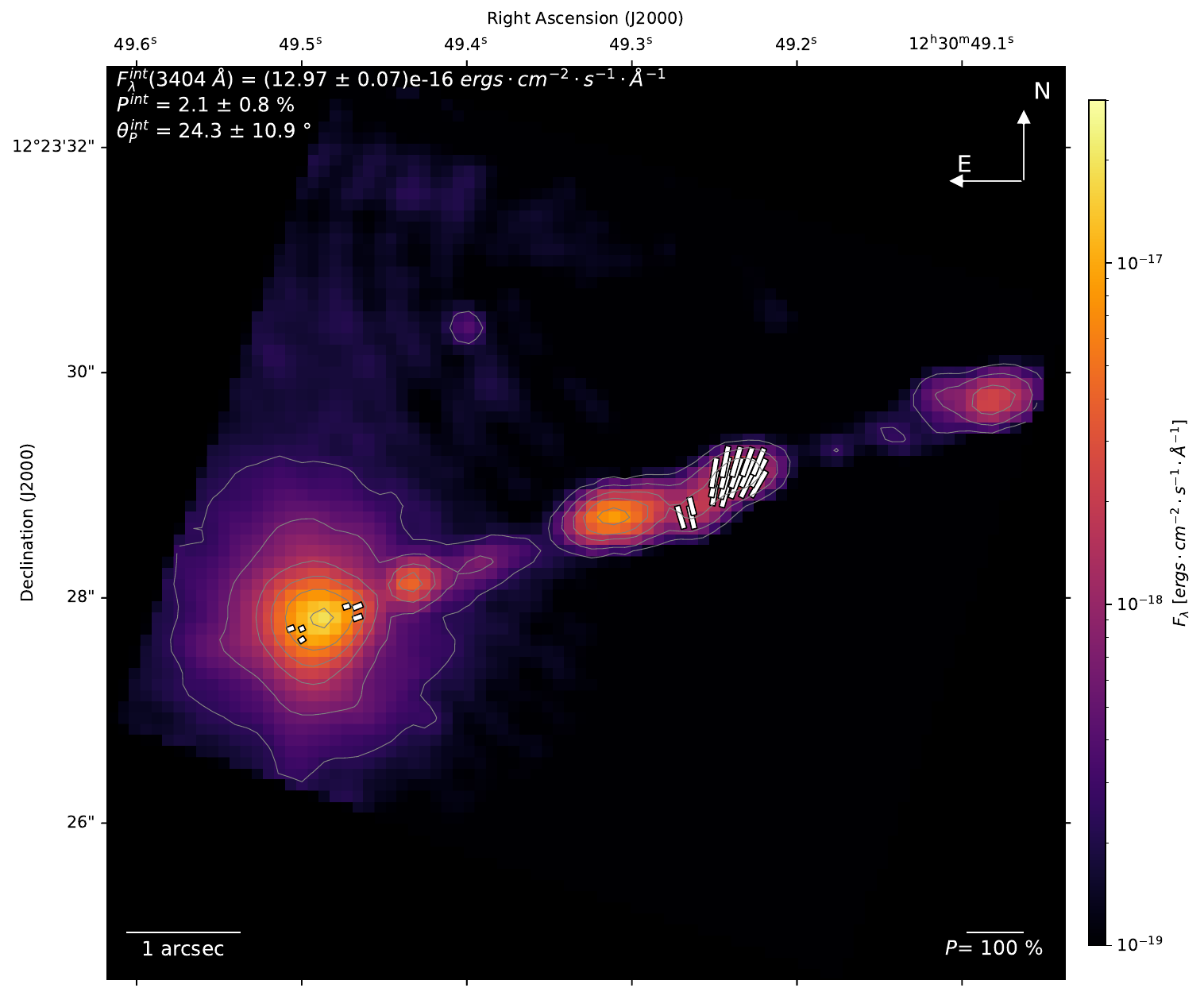}
\caption{Same as Fig.~\ref{Fig:POS~1_maps_1995}, but for the fourth observation (1998).}
\label{Fig:POS~1_maps_1998}
\end{figure*}

\begin{figure*}[htp]
\centering
\includegraphics[width=\textwidth]{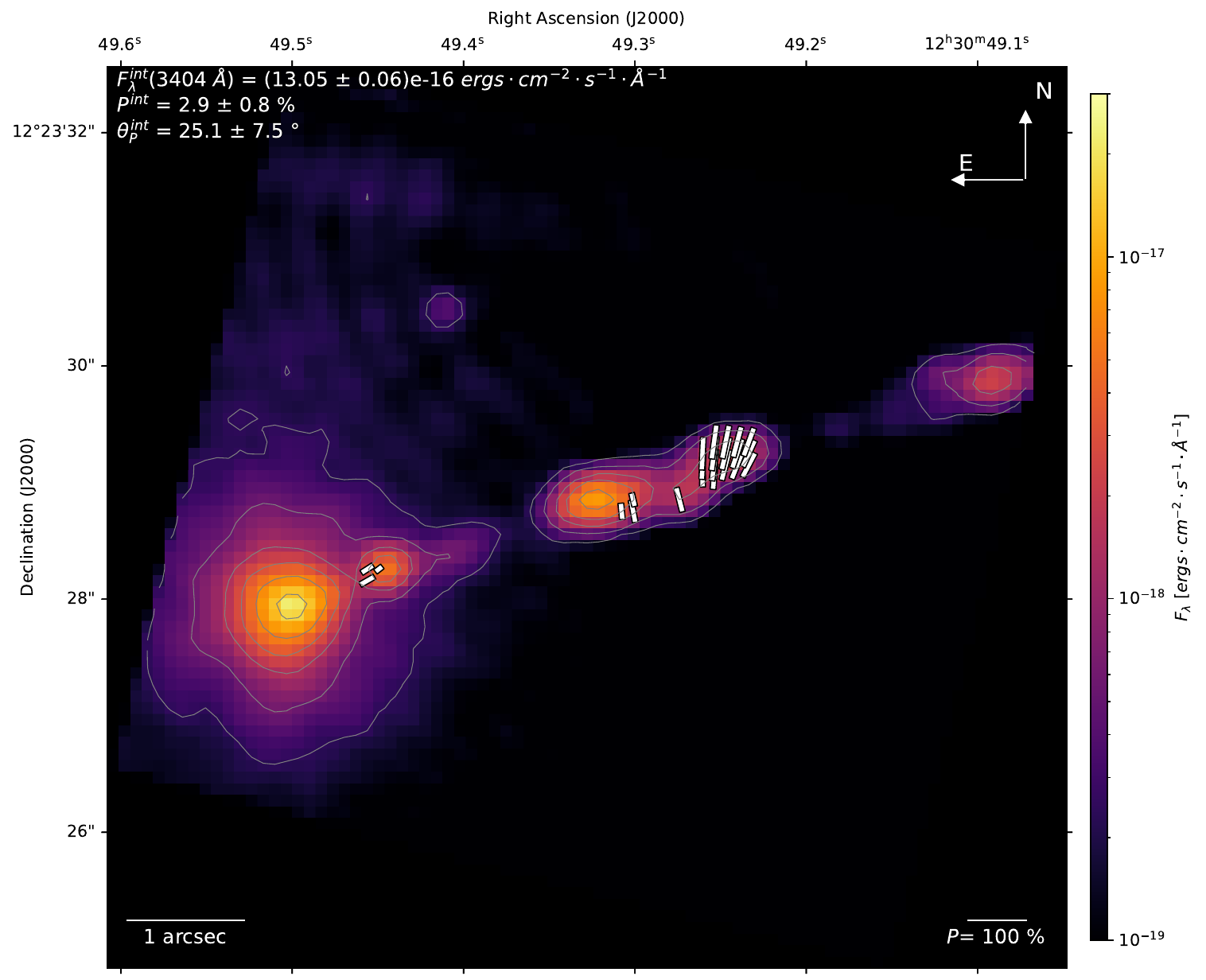}
\caption{Same as Fig.~\ref{Fig:POS~1_maps_1995}, but for the fifth observation (1999).}
\label{Fig:POS~1_maps_1999}
\end{figure*}

In the following, we show the large scale maps of the POS~1 section of the jet, from 1995 to 1999 (Figs.~\ref{Fig:POS~1_maps_1995}, \ref{Fig:POS~1_maps_1996}, \ref{Fig:POS~1_maps_1997}, \ref{Fig:POS~1_maps_1998} and \ref{Fig:POS~1_maps_1999}). As already mentioned in the main text, the maps have been resampled to pixels of $0.1'' \times 0.1''$. Intensity is colour-coded in erg cm$^{-2}$ s$^{-1}$ \AA$^{-1}$, from 10$^{-19}$ to 3 $\times$ 10$^{-17}$. The contours are displayed for 0.8\%, 2\%, 5\%, 10\%, 20\% and 50\% of the maximum flux. The polarization vectors are displayed in white for $\left[\text{S/N}\right]_P \geq 3$. On the top-left corner of the maps are displayed the flux and polarization values, integrated over the whole FOC FoV ($7'' \times 7''$).

\section{polarized maps of POS~3}
\label{AppendixB}

\begin{figure*}[htp]
\centering
\includegraphics[width=\textwidth]{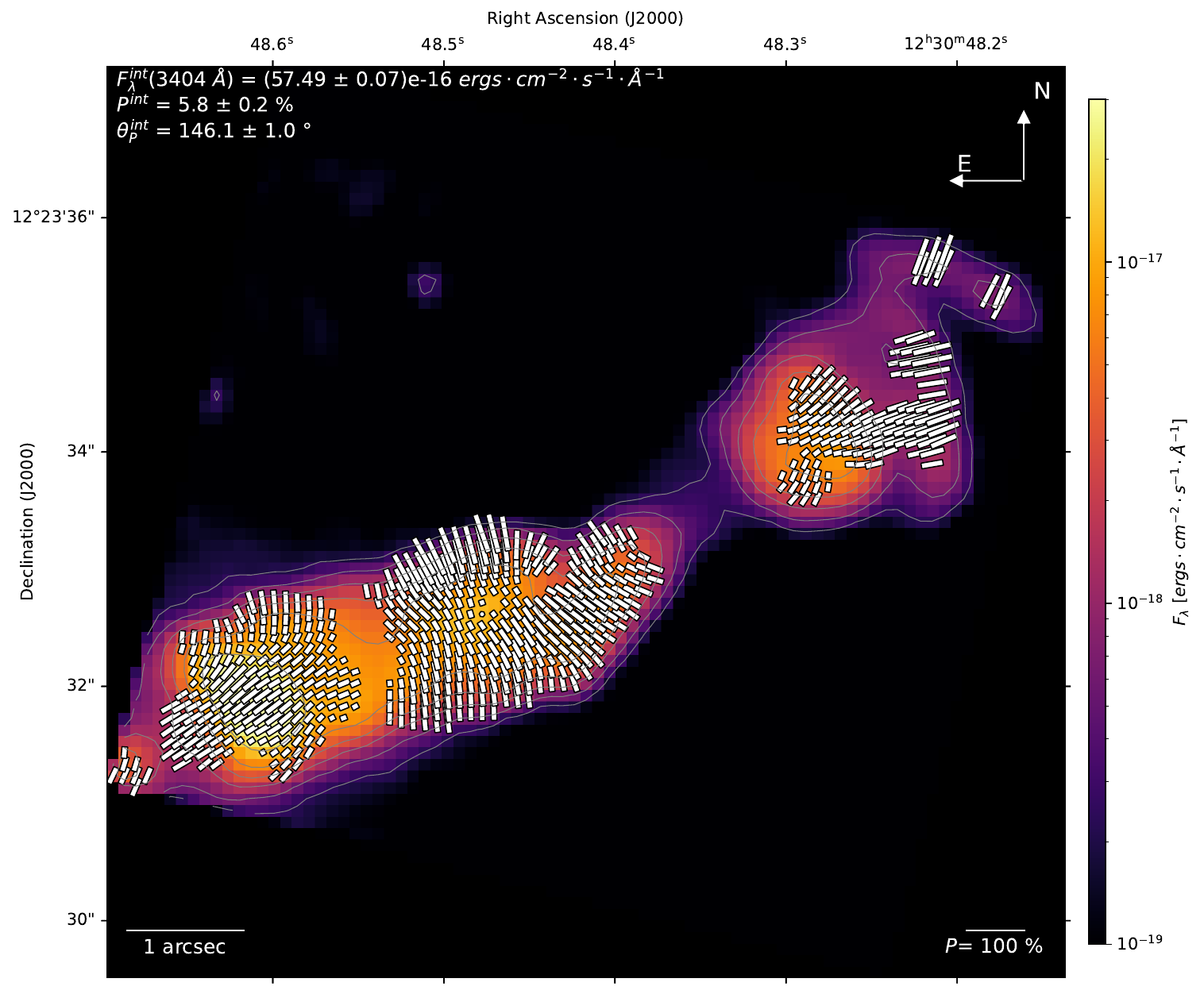}
\caption{HST/FOC 1995 observation of M87's POS~3 resampled to pixels of $0.1'' \times 0.1''$. Intensity is colour-coded in erg cm$^{-2}$ s$^{-1}$ \AA$^{-1}$. polarization vectors are displayed for $\left[\text{S/N}\right]_P \geq 3$. The contours are displayed for 0.8\%, 2\%, 5\%, 10\%, 20\% and 50\% of the maximum flux. On the top-left corner of the maps are displayed the flux and polarization values, integrated over the whole FOC FoV ($7'' \times 7''$).}
\label{Fig:POS~3_maps_1995}
\end{figure*}

\begin{figure*}[htp]
\centering
\includegraphics[width=\textwidth]{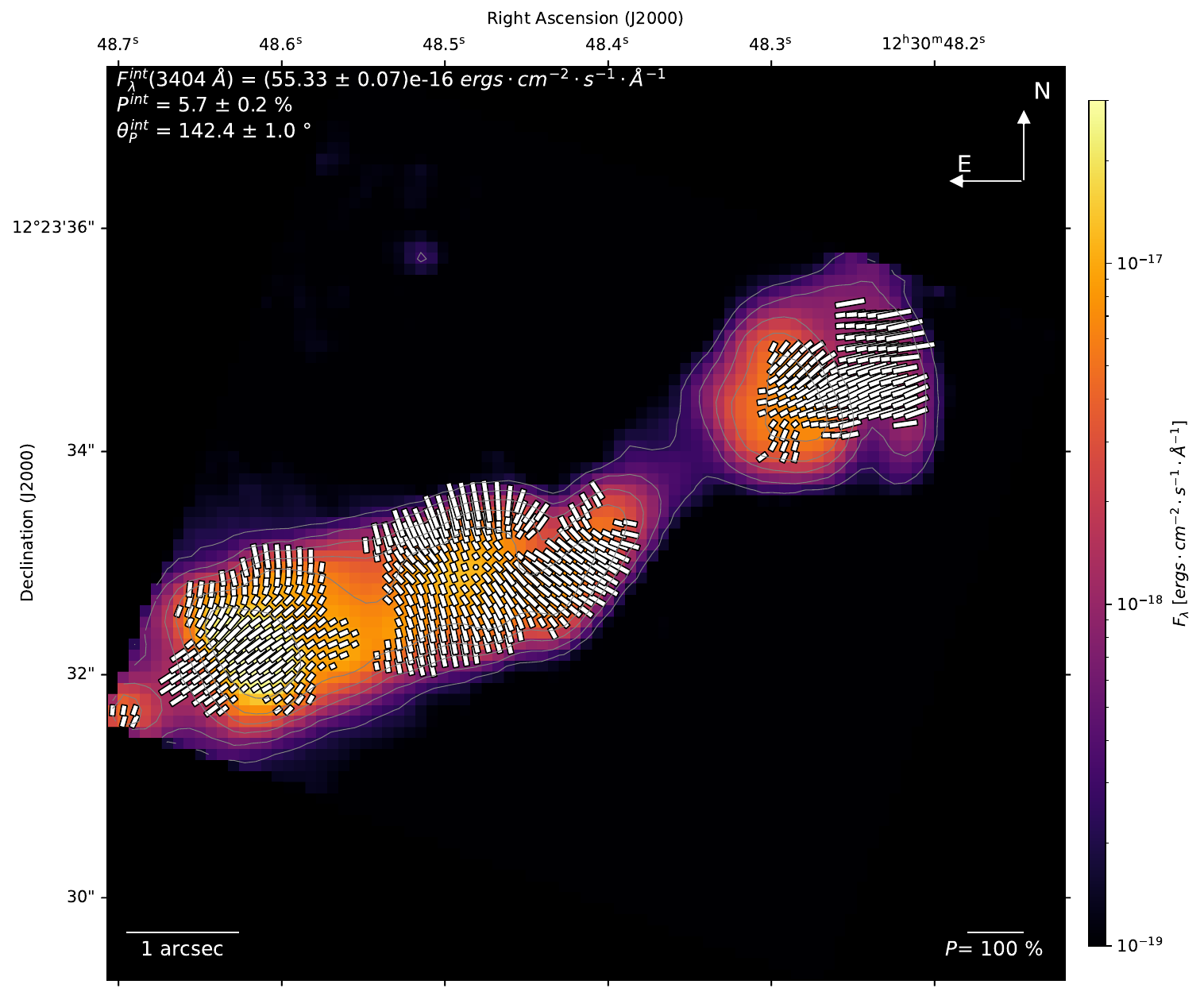}
\caption{Same as Fig.~\ref{Fig:POS~3_maps_1995}, but for the second observation (1996).}
\label{Fig:POS~3_maps_1996}
\end{figure*}

\begin{figure*}[htp]
\centering
\includegraphics[width=\textwidth]{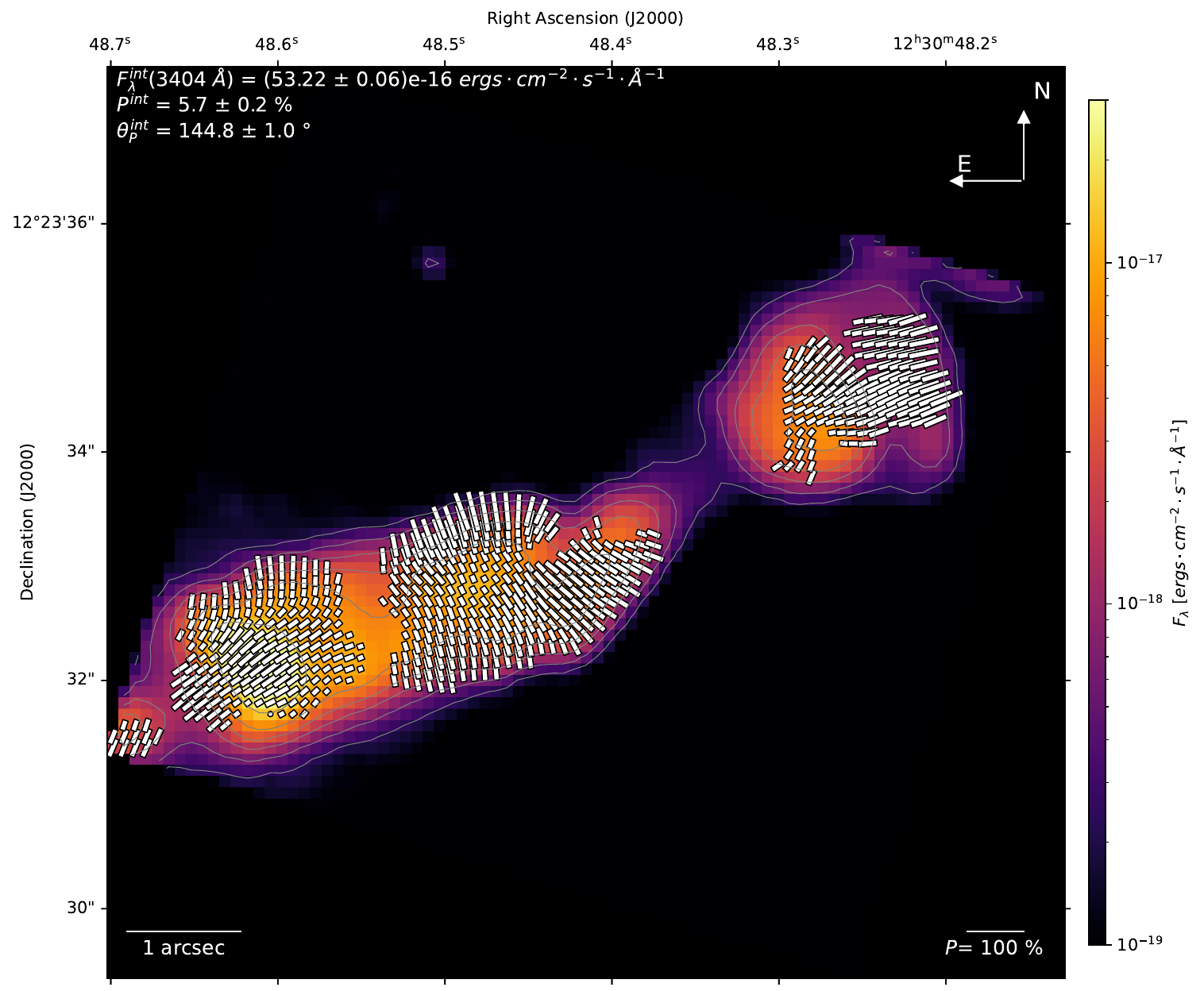}
\caption{Same as Fig.~\ref{Fig:POS~3_maps_1995}, but for the third observation (1997).}
\label{Fig:POS~3_maps_1997}
\end{figure*}

\begin{figure*}[htp]
\centering
\includegraphics[width=\textwidth]{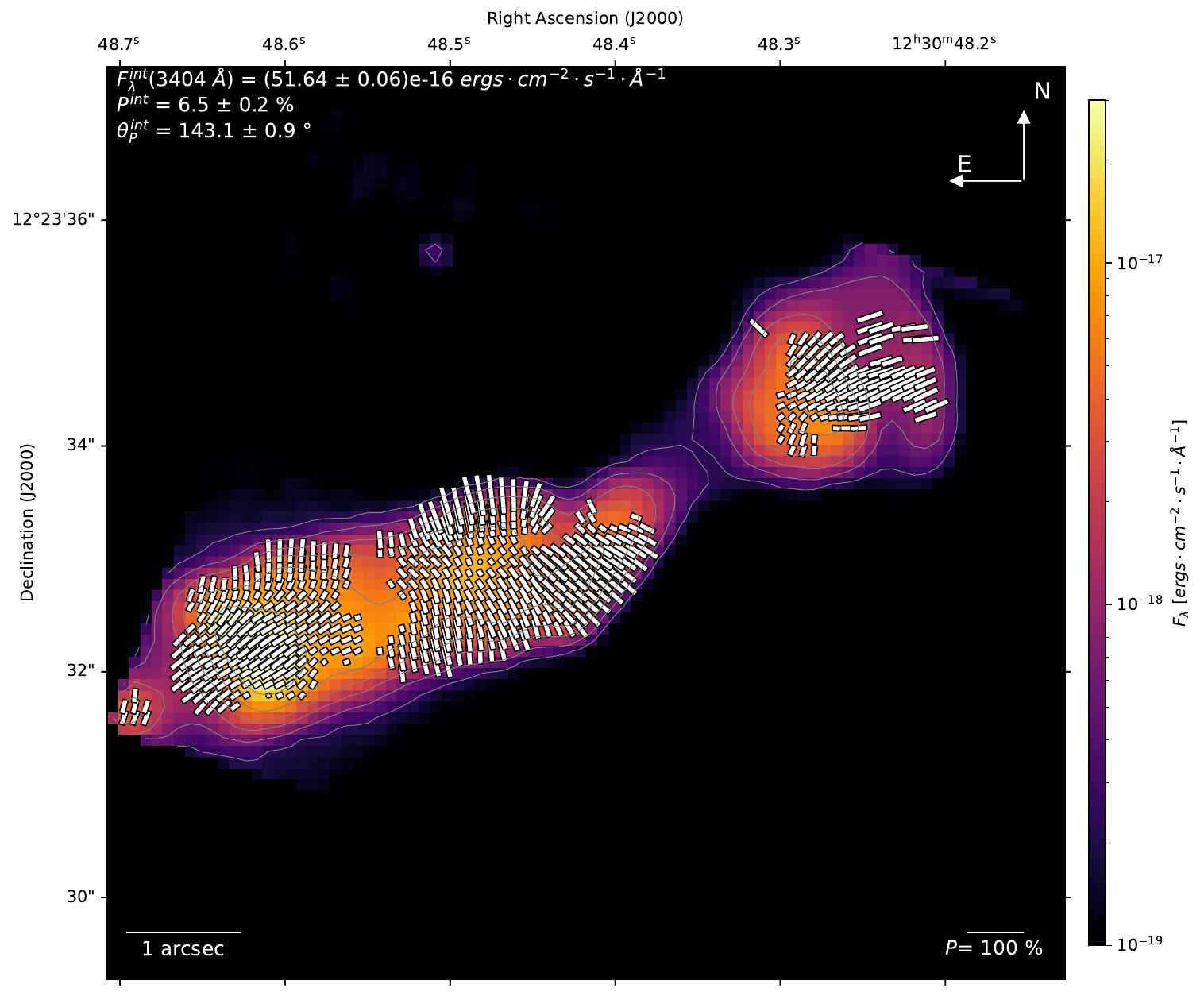}
\caption{Same as Fig.~\ref{Fig:POS~3_maps_1995}, but for the fourth observation (1998).}
\label{Fig:POS~3_maps_1998}
\end{figure*}

\begin{figure*}[htp]
\centering
\includegraphics[width=\textwidth]{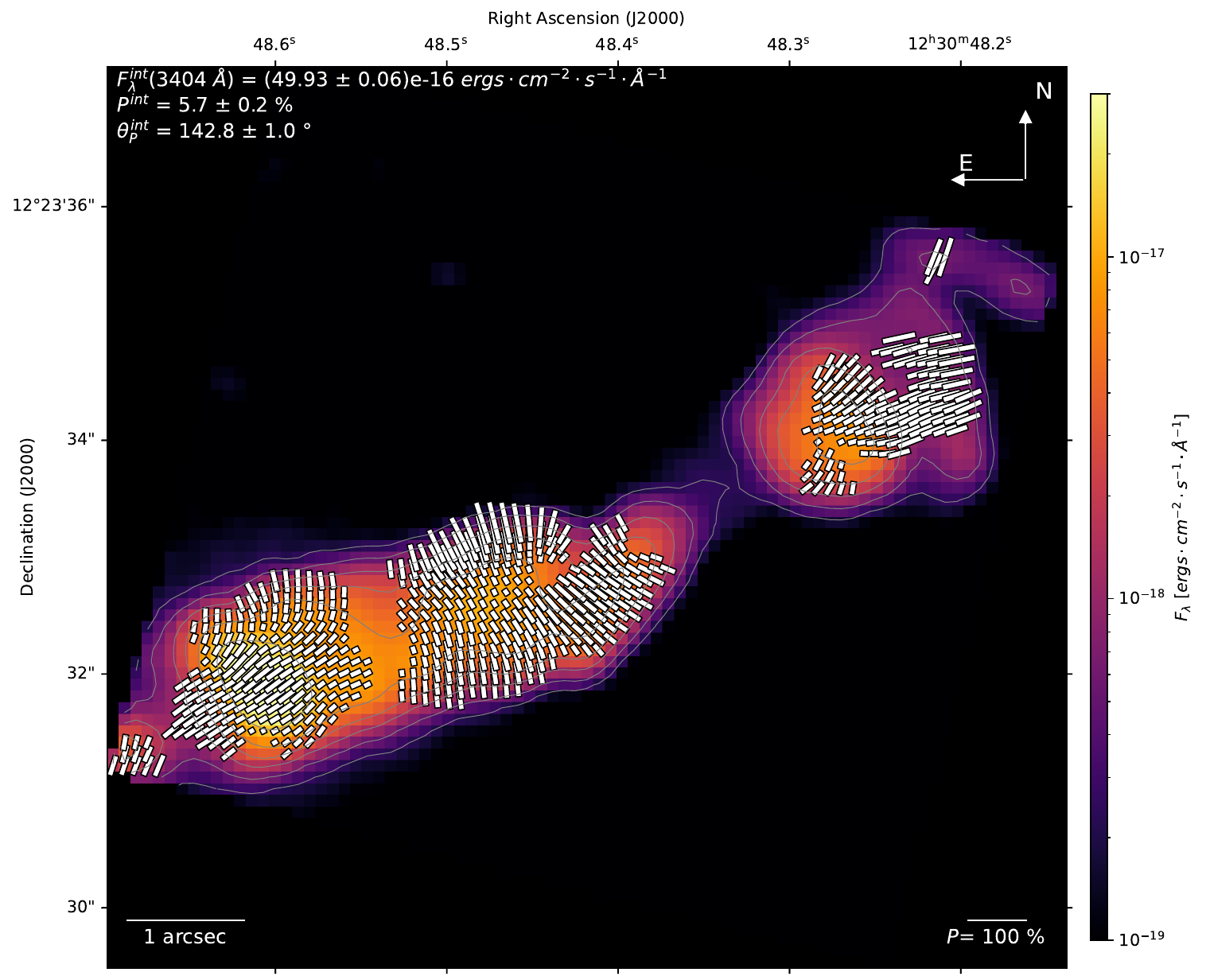}
\caption{Same as Fig.~\ref{Fig:POS~3_maps_1995}, but for the fifth observation (1999).}
\label{Fig:POS~3_maps_1999}
\end{figure*}

In the following, we show the large scale maps of the POS~3 section of the jet, from 1995 to 1999 (Figs.~\ref{Fig:POS~3_maps_1995}, \ref{Fig:POS~3_maps_1996}, \ref{Fig:POS~3_maps_1997}, \ref{Fig:POS~3_maps_1998} and \ref{Fig:POS~3_maps_1999}). The details are the same as in Sect.~\ref{AppendixA}.

\end{appendix}

\end{document}